\documentclass[lettersize,journal]{IEEEtran}
\usepackage{amsmath,amsfonts}
\usepackage{amssymb}
\usepackage{algorithm}
\usepackage{array}
\usepackage[font={footnotesize}]{caption}
\usepackage{subcaption}
\usepackage{textcomp}
\usepackage{stfloats}
\usepackage{url}
\usepackage{verbatim}
\usepackage{balance}
\usepackage{graphicx}
\usepackage{cite}
\usepackage{mathtools}
\usepackage[nolist]{acronym}
\usepackage{bm}
\usepackage[mode=image]{standalone}
\usepackage{tikz}
\usepackage{multirow}
\usepackage{pgfplots}
\usepackage{bigfoot} 
\usepackage{booktabs}
\usepackage{ragged2e}
\usepackage{calc} 
\usepackage[multiple]{footmisc}

\usetikzlibrary{arrows.meta}
\tikzset{every picture/.style={line width=0.6pt}}
\usetikzlibrary{arrows.meta,
                calc, 
                backgrounds,
                chains,
                fit,
                quotes,
                shapes.geometric,
                positioning,
                intersections,
                spath3,
                spy}

\usepgfplotslibrary{colormaps} 
\usepgfplotslibrary{fillbetween}
\usetikzlibrary{pgfplots.colormaps} 

\usepackage{algpseudocode}

\algdef{SE}[SUBALG]{Indent}{EndIndent}{}{\algorithmicend\ }%
\algtext*{Indent}
\algtext*{EndIndent}

\newtheorem{example}{Example}
\newtheorem{remark}{Remark}

\usepackage{threeparttable}

\newcommand{\hermit}{\mathsf{H}}

\newcommand{\bmax}{\bm{a}_{\mathrm{x}}}
\newcommand{\bmatx}{\bm{a}_{\mathrm{tx}}}
\newcommand{\bmarx}{\bm{a}_{\mathrm{rx}}}

\newcommand{\bmf}{\bm{f}}
\newcommand{\bmfc}{\bm{f}_{\mathrm{c}}}
\newcommand{\bmfs}{\bm{f}_{\mathrm{s}}}
\newcommand{\bmfx}{\bm{f}_{\mathrm{x}}}

\newcommand{\bmfbar}{\bar{\bm{f}}}
\newcommand{\bmp}{\bm{p}}
\newcommand{\bmphat}{\hat{\bm{p}}}
\newcommand{\bmgammax}{\bm{\gamma}_{\mathrm{x}}}
\newcommand{\bmgammatx}{\bm{\gamma}_{\mathrm{tx}}}
\newcommand{\bmgammarx}{\bm{\gamma}_{\mathrm{rx}}}

\newcommand{\bmpx}{\bm{p}_{\mathrm{x}}}
\newcommand{\bmptx}{\bm{p}_{\mathrm{tx}}}
\newcommand{\bmprx}{\bm{p}_{\mathrm{rx}}}
\newcommand{\bmu}{\bm{u}}

\newcommand{\bmx}{\bm{x}}

\newcommand{\bmrho}{\bm{\rho}}
\newcommand{\bmS}{\bm{S}}
\newcommand{\bmY}{\bm{Y}}
\newcommand{\bmW}{\bm{W}}
\newcommand{\bmw}{\bm{w}}
\newcommand{\bmkappa}{\bm{\kappa}}
\newcommand{\bmXi}{\bm{\Xi}}
\newcommand{\bmXitx}{\bm{\Xi}_{\mathrm{tx}}}
\newcommand{\bmXirx}{\bm{\Xi}_{\mathrm{rx}}}

\DeclareDocumentCommand{\complexset}{o o}{%
    \mathbb{C}\IfValueT{#1}{\IfValueTF{#2}{^{#1\times#2}}{^{#1}}}
}
\DeclareDocumentCommand{\realset}{o o}{%
    \mathbb{R}\IfValueT{#1}{\IfValueTF{#2}{^{#1\times#2}}{^{#1}}}
}

\newcommand{\cnormal}{\mathcal{CN}}

\newcommand{\Deltaf}{\Delta_{\mathrm{f}}}
\renewcommand{\d}{\mathrm{d}}

\newcommand{\E}{\mathbb{E}}
\newcommand{\Exp}{\mathrm{Exp}}

\newcommand{\Lcal}{\mathcal{L}}

\newcommand{\norm}[1]{\lVert#1\rVert}
\newcommand{\normfro}[1]{\lVert#1\rVert_{\mathrm{F}}}

\newcommand{\sigmarcst}{\sigma_{\mathrm{rcs}, t}}

\newcommand{\Rt}{R_t}
\newcommand{\Rmin}{R_{\min}}
\newcommand{\Rmax}{R_{\max}}
\newcommand{\taumin}{\tau_{\min}}
\newcommand{\taumax}{\tau_{\max}}

\newcommand{\Tmax}{T_{\max}}
\newcommand{\thetamin}{\theta_{\min}}
\newcommand{\thetamax}{\theta_{\max}}

\newcommand{\U}{\mathcal{U}}

\newcommand{\Xset}{\mathcal{X}}

\DeclareDocumentCommand{\Ys}{o}{\bmY_{\mathrm{s}\IfValueT{#1}{,#1}}}
\DeclareDocumentCommand{\yc}{o}{%
  \bm{\mathrm{y}}_{\mathrm{c}\IfValueT{#1}{,#1}}%
}

\newcommand{\alphatilde}{\Tilde{\alpha}}
\newcommand{\Rtilde}{\Tilde{R}}
\newcommand{\sigmarcstt}{\Tilde{\sigma}_{\mathrm{rcs},t}}
\newcommand{\Ttilde}{\Tilde{T}}
\newcommand{\Ttildemax}{\Tilde{T}_{\max}}
\newcommand{\thetatilde}{\Tilde{\theta}}
\newcommand{\tautilde}{\Tilde{\tau}}
\newcommand{\thetatmin}{\Tilde{\theta}_{\min}}
\newcommand{\thetatmax}{\Tilde{\theta}_{\max}}
\newcommand{\thetabmin}{\bar{\theta}_{\min}}
\newcommand{\thetabmax}{\bar{\theta}_{\max}}
\newcommand{\thetafov}{\theta_{\mathrm{fov}}}

\newcommand{\omegar}{\omega_{\mathrm{r}}}

\newcommand{\etar}{\eta_{\mathrm{r}}}

\newcommand{\Ntheta}{N_{\mathrm{\theta}}}
\newcommand{\Ntau}{N_{\mathrm{\tau}}}
\newcommand{\bartheta}{\bar{\theta}}

\newcommand{\bmPhia}{\bm{\Phi}_{\mathrm{a}}}
\newcommand{\bmPhid}{\bm{\Phi}_{\mathrm{d}}}

\newcommand{\Pset}{\mathcal{P}}
\newcommand{\Psethat}{\hat{\mathcal{P}}}
\newcommand{\thetahat}{\hat{\theta}}
\newcommand{\tauhat}{\hat{\tau}}

\newcommand{\bmL}{\bm{L}}

\newcommand{\bmalphahat}{\hat{\bm{\alpha}}}
\newcommand{\ihat}{\hat{i}}
\newcommand{\jhat}{\hat{j}}

\newcommand{\Lr}{\mathcal{L}_{\mathrm{r}}}
\newcommand{\Lc}{\mathcal{L}_{\mathrm{c}}}
\newcommand{\Lbar}{\bar{\mathcal{L}}}

\newcommand{\thetaint}{\bm{\theta}_{\mathrm{int}}}
\newcommand{\bmzeta}{\bm{\zeta}}

\newcommand{\pmd}{p_{\mathrm{md}}}
\newcommand{\pfa}{p_{\mathrm{fa}}}
\newcommand{\That}{\hat{T}}

\newcommand{\bmxenc}{\bmx_{\mathrm{enc}}}
\newcommand{\bmxref}{\bmx_{\mathrm{ref}}}

\newcommand{\degreee}{^{\circ}}
\newcommand{\normal}{\mathcal{N}}
\newcommand{\sigmarcsmean}{\sigma_{\mathrm{mean}}}
\newcommand{\thetamean}{\theta_{\mathrm{mean}}}
\newcommand{\Deltatheta}{\Delta_{\theta}}
\newcommand{\SCNRs}{\mathrm{SCNR}_{\mathrm{s}}}
\newcommand{\SNRc}{\mathrm{SNR}_{\mathrm{c}}}
\newcommand{\Expectation}{\mathbb{E}}
\newcommand{\bmpideal}{\bm{p}_{\mathrm{ideal}}}
\newcommand{\bmvarepsilonp}{\bm{\varepsilon}_{\mathrm{p}}}
\newcommand{\bmvarepsilonf}{\bm{\varepsilon}_{\mathrm{f}}}
\newcommand{\Tcp}{T_{\mathrm{cp}}}
\newcommand{\dcut}{d^{(\gamma)}}
\newcommand{\abs}[1]{|#1|}

\newcommand{\bmfpert}{\tilde{\bm{f}}}

\newcommand{\bmPsi}{\bm{\Psi}}
\newcommand{\bmPsitx}{\bm{\Psi}_{\mathrm{tx}}}
\newcommand{\bmPsirx}{\bm{\Psi}_{\mathrm{rx}}}
\newcommand{\bmxitx}{\bm{\xi}_{\mathrm{tx}}}
\newcommand{\bmomegatx}{\bm{\omega}_{\mathrm{tx}}}
\newcommand{\bmxirx}{\bm{\xi}_{\mathrm{rx}}}
\newcommand{\bmomegarx}{\bm{\omega}_{\mathrm{rx}}}
\newcommand{\xitx}{{\xi}_{\mathrm{tx}}}
\newcommand{\omegatx}{{\omega}_{\mathrm{tx}}}
\newcommand{\xirx}{{\xi}_{\mathrm{rx}}}
\newcommand{\omegarx}{{\omega}_{\mathrm{rx}}}

\DeclareDocumentCommand{\xix}{o}{%
  \ensuremath{\xi_{\mathrm{x\IfValueT{#1}{,#1}}}}%
}

\DeclareDocumentCommand{\omegax}{o}{%
  \ensuremath{\omega_{\mathrm{x\IfValueT{#1}{,#1}}}}%
}

\newcommand{\bmchihat}{\hat{\bm{\chi}}}

\newcommand{\Cmax}{C_{\max}}
\newcommand{\sigmac}{\sigma_{\mathrm{c}}}

\newcommand{\bigO}{\mathcal{O}}

\newcommand{\thetap}{\theta_{\mathrm{p}}}
\newcommand{\sigmatheta}{\sigma_{\theta}}

\newcommand{\Psuper}{\textsuperscript{\P}}

\newcommand{\Iset}{\mathcal{I}}
\newcommand{\Isettilde}{\tilde{\mathcal{I}}}

\newcommand{\bmAa}{\bm{A}_{\mathrm{a}}}
\newcommand{\bmAd}{\bm{A}_{\mathrm{d}}}

\begin{acronym}[ACRONYM]
\acro{6G}{6th generation wireless systems}
\acro{ADI}{antenna displacement impairment}
\acro{ADM}{angle-delay map}
\acro{AWGN}{additive white Gaussian noise}
\acro{BS}{base station}
\acro{CCE}{categorical cross-entropy}
\acro{CP}{cyclic prefix}
\acro{CSI}{channel state information}
\acro{DL}{deep learning}
\acro{DoA}{direction of arrival}
\acro{DoD}{direction of departure}
\acro{E2E}{end-to-end}
\acro{FOV}{field of view}
\acro{GF}{gradient-free}
\acro{GOSPA}{generalized optimal sub-pattern assignment}
\acro{GPI}{gain-phase impairment}
\acro{ISAC}{integrated sensing and communication}
\acro{LMMSE}{linear minimum mean-squared-error}
\acro{LoS}{line-of-sight}
\acro{NLOS}{non-line-of-sight}
\acro{NN}{neural network}
\acro{MB-ML}{model-based machine learning}
\acro{MIMO}{multiple-input multiple-output}
\acro{MLE}{maximum likelihood estimation}
\acro{MUSIC}{multiple signal classification}
\acro{OFDM}{orthogonal frequency-division multiplexing}
\acro{PMF}{probability mass function}
\acro{PSK}{phase-shift keying}
\acro{QAM}{quadrature amplitude modulation}
\acro{OMP}{orthogonal matching pursuit}
\acro{PDF}{probability density function}
\acro{POGD}{projected online gradient descent}
\acro{RCS}{radar cross section}
\acro{RL}{reinforcement learning}
\acro{RX}{receiver}
\acro{SER}{symbol error rate}
\acro{SIMO}{single-input multiple-output}
\acro{SL}{supervised learning}
\acro{SLCB}{supervised learning with channel backpropagation}
\acro{SSL}{self-supervised learning}
\acro{SCNR}{signal-to-clutter-plus-noise ratio}
\acro{SNR}{signal-to-noise ratio}
\acro{TX}{transmitter}
\acro{TRX}{transceiver}
\acro{UE}{user equipment}
\acro{UL}{unsupervised learning}
\acro{ULA}{uniform linear array}
\end{acronym}

\usepackage[all=normal,paragraphs=tight,floats=normal,mathspacing=normal,wordspacing=tight,charwidths=tight,mathdisplays=normal,leading=normal]{savetrees}

\begin{document}

\title{Unsupervised End-to-End Array Calibration for Multi-Target Integrated Sensing and Communication}

\author{Jos\'{e} Miguel Mateos-Ramos,~\IEEEmembership{Student Member,~IEEE}, 
Baptiste Chatelier,~\IEEEmembership{Member,~IEEE}
 \\
Luc Le Magoarou,~\IEEEmembership{Member,~IEEE}, 
Nir Shlezinger,~\IEEEmembership{Senior Member,~IEEE}, \\
Henk Wymeersch,~\IEEEmembership{Fellow,~IEEE},
Christian H\"{a}ger,~\IEEEmembership{Member,~IEEE}

\thanks{
This work was supported, in part, by a grant from the Chalmers AI Research Center Consortium (CHAIR), the Swedish Foundation for Strategic Research (SSF) (grant FUS21-0004, SAICOM), and Swedish Research Council (VR grant 2022-03007). The computations were enabled by resources provided by the National Academic Infrastructure for Supercomputing in Sweden (NAISS), partially funded by the Swedish Research Council through grant agreement no. 2022-06725. The work of C.~Häger was also supported by the Swedish Research Council under grant no. 2020-04718.}
\thanks{Jos\'{e} Miguel Mateos-Ramos, Henk Wymeersch, and Christian H\"{a}ger are with the Department of Electrical Engineering, Chalmers University of Technology, Sweden (email: josemi@chalmers.se; henkw@chalmers.se; christian.haeger@chalmers.se).}%
\thanks{Baptiste Chatelier and Luc Le Magoarou are with INSA Rennes, CNRS, IETR-UMR 6164, F-35000, Rennes, France (email: Luc.Le-Magoarou@insa-rennes.fr).}%
\thanks{Nir Shlezinger is with the School of ECE, Ben-Gurion University of the Negev, Be'er Sheva 8410501, Israel (email: nirshl@bgu.ac.il).}
}



\maketitle
\begin{abstract}
    In this work, we consider end-to-end calibration of an integrated sensing and communication (ISAC) base station (BS) under gain-phase and antenna displacement impairments without collecting signals from predefined positions (labeled data). We consider a BS with two impaired uniform linear arrays used for simultaneous multi-target sensing and communication with a user equipment (UE) leveraging orthogonal frequency-division multiplexing signals. 
    The main contribution is the design of a framework that can compensate for the impairments without labeled data and considering coherent receive signals.
    We harness a differentiable precoder based on the maximum array response in an angular direction at the transmitter and the orthogonal matching pursuit (OMP) algorithm at the sensing receiver. 
    We propose an ISAC loss as a combination of sensing and communication losses that provides a trade-off between the two functionalities. We compare two sensing objective alternatives: (i) maximize the maximum response of the angle-delay map of the targets or (ii) minimize the norm of the residual signal at the output of the OMP algorithm after all estimated targets have been removed. The communication objective maximizes the energy of the received signal at the UE. 
    Additionally, our framework leverages an approximation of the channel gradient that avoids the impractical knowledge of the gradient of the channel. 
    Our results show that the proposed method performs closely to using labeled data and knowledge of the channel gradient in terms of sensing position estimation and communication symbol error rate. When comparing the two sensing losses, minimizing the norm of the OMP residual yields significantly better sensing position estimation with slightly increased complexity.
\end{abstract}


\section{Introduction}
\label{sec:intro}
\Ac{ISAC} combines communication and sensing capabilities to mutually benefit each other and efficiently use wireless resources. \Ac{ISAC} is considered a key pillar of the forthcoming \ac{6G} standard~\cite{liu2022integrated}. It offers improved hardware, energy, and spectral efficiency compared to dedicated sensing and communication systems~\cite{cui2023integrated, wei2023integrated}. The benefits of ISAC enable new \ac{6G} applications such as vehicle-to-everything communications, human activity sensing, and
unmanned aerial vehicle networks~\cite{lu2024integrated}.

Conventional \ac{ISAC} techniques are largely based on physical and mathematical models of the transmitted and received waveforms, which are used to design the corresponding \acp{TX} and \acp{RX}~\cite{he2022joint, zhao2022radio, wen2023efficient, wei2023iterative}. However, hardware impairments introduce calibration errors that lead to model mismatches, resulting in degraded sensing and communication performance~\cite{koivunen2024multicarrier}. This issue becomes particularly critical in \ac{6G} systems, where \acp{BS} are expected to employ large-scale antenna arrays to enhance communication capacity and sensing angular resolution. In such multi-antenna deployments,  array calibration errors can significantly distort the effective array response, directly impacting both functionalities. Accordingly, in this work we focus on the calibration of antenna arrays for multi-antenna \ac{ISAC} \acp{BS}.

\subsection*{Calibration Methods} \label{subsec:cal_methods}
Model-based calibration relies on mathematical models of the signal propagation in the environment to compensate for impairments. We classify the model-based calibration literature into: $(i)$ in-chamber calibration, $(ii)$ in-situ calibration, and $(iii)$ self-calibration.
The most traditional method consists of calibrating an antenna in an anechoic chamber (in-chamber calibration). 
An anechoic chamber provides a controlled environment to perform calibration based on \ac{LoS} propagation~\cite{vassanelli2020calibration, yang2022theory, pan2022efficient}. In this environment, the \ac{LoS} models, and hence the calibration algorithms, are more accurate than in the operating environment. However, residual calibration errors may exist during deployment of the antenna in the real scenario due to installation errors, cable deformations, or changes in the coupling due to different scattering in the array's environment~\cite{gupta2003experimental, mubarak2013characterizing, sippel2020insitu}. Additionally, calibration in an anechoic chamber is expensive and time-consuming. 

Alternative model-based methods perform in-situ calibration by collecting measurements in the actual operating environment at known positions~\cite{pan2023insitu, sippel2020insitu}. In-situ calibration is suitable for environments where the positions of the signal sources or the targets are known and it can compensate for the impairments involved during deployment, e.g., installation errors. 
However, in a real \ac{ISAC} environment, gathering data from sensing objects at known positions may be impractical or expensive.

The third framework is self-calibration, which seeks to jointly estimate target parameters and array impairments directly from online measurements, without requiring signals from known positions~\cite{liu2013unified, wang2019doa, chen2020newatomic, wang2024robust, li2020joint, li2013joint, li2015transmit, guo2017angle}. 
Self-calibration is particularly attractive in dynamic sensing environments, where deploying calibration sources or collecting measurements at predefined locations is infeasible. By exploiting structural properties of the received signals, these methods aim to disentangle propagation parameters and hardware impairments in a fully data-driven manner, enabling autonomous operation after deployment.
Representative works adopt sparse or structured signal models to achieve this goal. In~\cite{liu2013unified}, array calibration and \ac{DoA} estimation are jointly performed using a sparse representation framework, where \acp{ADI} are modeled via a Taylor expansion and estimated together with the \ac{DoA} parameters using an expectation-maximization procedure; extensions also account for mutual coupling and \acp{GPI}. 
The work in~\cite{wang2019doa} addresses mutual coupling through sparse recovery over an over-complete \ac{DoA} grid, relaxing the resulting $\ell^0$-norm problem into an $\ell^1$-regularized formulation. 
A gridless approach based on atomic norm minimization under \acp{GPI} is proposed in~\cite{chen2020newatomic}, while~\cite{wang2024robust} develops a low-rank row-sparse covariance decomposition method for calibration under \acp{GPI}. 
Self-calibration has also been studied in \ac{MIMO} radar, where the \ac{TX} and \ac{RX} arrays are impaired simultaneously and their errors appear coupled in the virtual array. For monostatic arrays,~\cite{li2020joint} refines a subspace-based coarse \ac{DoA} estimate through a local search of an improved \ac{MUSIC} cost function and obtains the \acp{GPI} from the noise subspace. For bistatic arrays, the \ac{TX} and \ac{RX} direction matrices are estimated by trilinear decomposition~\cite{li2013joint} or by an ESPRIT-based algorithm~\cite{li2015transmit}, and the \acp{GPI} are subsequently recovered in closed form, while~\cite{guo2017angle} groups mutual coupling, \acp{GPI}, and \acp{ADI} into a single angularly dependent gain-phase error that is later decomposed into the three impairment types by nonlinear least squares. In all of the MIMO radar works, the \ac{TX} array radiates mutually orthogonal waveforms, so that the
contribution of each \ac{TX} element is separated at the \ac{RX} by a bank of matched filters and
both arrays are jointly observed through a virtual array.

Despite their effectiveness, these methods are tailored to specific settings: (i) several assume noncoherent sources~\cite{liu2013unified, wang2019doa, wang2024robust, li2020joint, li2013joint, li2015transmit, guo2017angle}, which limits their applicability in single-snapshot monostatic sensing where reflections cannot decorrelate over slow-time; (ii) others focus on a single impairment type (e.g., mutual coupling or \acp{GPI})~\cite{wang2019doa, chen2020newatomic, wang2024robust, li2020joint, li2013joint, li2015transmit}; (iii)  the \ac{MIMO} radar works in~\cite{li2020joint, li2013joint, li2015transmit,
guo2017angle} calibrate the \ac{TX} and \ac{RX} arrays jointly, but they recover the \ac{TX}
impairments from the virtual array enabled by orthogonal waveforms, which is unavailable when the
transmitted waveform is dictated by the communication service; and (iv) most are developed for single-carrier systems and multiple snapshots, leaving multi-carrier \ac{ISAC} scenarios largely unexplored and increasing estimation and calibration delays.

An alternative paradigm to model-based calibration aims at learning to calibrate in a data-driven fashion. The main data-driven approaches for calibration can be classified as purely \ac{DL} or as a form of \ac{MB-ML}. Data-driven approaches offer more flexibility than model-based methods to adapt to modeling mismatches as the former do not rely on mathematical models and calibration is purely based on data. \Ac{DL} leverages \acp{NN} in a black-box manner to perform signal design or parameter estimation. 
\Ac{MB-ML} parameterizes existing model-based designs and algorithms while maintaining their computational graph as a blueprint~\cite{shlezinger2023deep}. \Ac{MB-ML} lies between model-based approaches and \ac{DL}, usually requiring less data and fewer training parameters and offering more explainability than \ac{DL}.
However, most data-driven approaches~\cite{famoriji2020intelligent, iye2022deep, gao2025nnassisted, chen2025novel,shmuel2025subspacenet, mateos2025model} require labeled data in the form of the true angle, angular spectrum, or position of the targets to perform calibration as a form of \ac{SL}. 

As opposed to \ac{SL}, some recent data-driven works perform \ac{UL}, which does not require labeled data for calibration~\cite{temiz2025unsupervised, chatelier2025physically, konstantino2026unsupervised, mateos2025unsupervised}. 
In~\cite{temiz2025unsupervised}, an \ac{NN} is used to compute the precoder of an ISAC \ac{BS} based on the estimated \ac{CSI}. 
The unsupervised loss function is rooted in the communication sum-rate and the sensing Cram\'{e}r-Rao lower bound of the \ac{DoD}, where the power constraint of the precoder is included as a penalty term in the designed loss function. 
The work in~\cite{chatelier2025physically} designs an \ac{MB-ML} differentiable version of the \ac{MUSIC} algorithm to perform \ac{DoA} estimation under \acp{GPI} and \acp{ADI}. A discrete grid of angles is considered and the steering matrix, parameterized by \acp{GPI} and \acp{ADI}, is iteratively refined by learning the physical impairments. 
The goal of the designed loss function is to maximize the \ac{MUSIC} spectrum around the estimated angles with the impaired antenna. 
In~\cite{konstantino2026unsupervised}, tracking of the \ac{DoA} over time is performed under \acp{ADI} and random perturbations of the \ac{RX} steering vector. An \ac{NN} is trained by assessing the deviation between the estimated \ac{DoA} of the \ac{NN} and the predicted \ac{DoA} based on a state evolution model of the \ac{DoA} over time. A loss function minimizing such deviation is proposed to calibrate the \ac{RX}.
The work in~\cite{mateos2025unsupervised} performs \ac{GPI} calibration at the sensing \ac{RX} in an \ac{ISAC} scenario. The position of a single target is estimated and an \ac{MB-ML} method is developed to compensate for the \acp{GPI} based on the response of the received signal to a discrete grid of angles and ranges.
The main limitations of~\cite{temiz2025unsupervised, chatelier2025physically, konstantino2026unsupervised, mateos2025unsupervised} are: $(i)$ \ac{TX} and \ac{RX} are individually optimized, but simultaneous calibration of both remains unexplored; $(ii)$ the results in~\cite{chatelier2025physically} show that \ac{UL} at the \ac{RX} side performs poorly compared to supervised learning; and $(iii)$ the work in \cite{konstantino2026unsupervised} leveraged downstream tracking to enable \ac{UL}, and is thus restricted to settings where sensing is coupled with subsequent target tracking. 

In view of the closest literature on model-based~\cite{liu2013unified, wang2019doa, chen2020newatomic, wang2024robust} and data-driven~\cite{temiz2025unsupervised, chatelier2025physically, mateos2025unsupervised,konstantino2026unsupervised} approaches for calibration, there is a need for an effective calibration method that can account for coherent signals and simultaneous \ac{TX} and \ac{RX} impairments without knowing the true positions of targets during calibration or the evolution of the targets over time. In Table~\ref{tab:literature}, we include a comparison of the closest literature with this work.

\begin{table*}[]
\centering%
\setlength{\tabcolsep}{2pt}%
\begin{threeparttable}
\caption{Comparison between this and closely related prior work. (MC: mutual coupling, NSV: noisy steering vector)}
\label{tab:literature}
\footnotesize
\begin{tabular}{@{}c||c|c|c|c|c|c|c|c|c@{}}
\toprule
Ref. & \begin{tabular}[c]{@{}c@{}}Calibration\\ type\end{tabular} & Impairments & Objective & Coherent signals & Multi-carrier & ISAC & \begin{tabular}[c]{@{}l@{}}TX or RX\\ calibration\end{tabular} & \begin{tabular}[c]{@{}c@{}}Monostatic /\\ bistatic\end{tabular} & \begin{tabular}[c]{@{}c@{}}Number of\\ snapshots\end{tabular} \\
\midrule \midrule
\cite{liu2013unified} & Model-based &  GPI, ADI, MC & DoA estimation & No & No & No & RX & N/A\makebox[0pt][l]{\Psuper} & 10--100 \\
\cite{wang2019doa} & Model-based & MC & DoA estimation & No & No& No & RX & N/A\makebox[0pt][l]{\Psuper} &  300 \\
\cite{chen2020newatomic} & Model-based & GPI & DoA estimation & Yes & No& No & RX & N/A\makebox[0pt][l]{\Psuper} & 5 \\
\cite{wang2024robust} & Model-based & GPI & DoA estimation & No & No& No & RX & N/A\makebox[0pt][l]{\Psuper} & 100 \\
\cite{li2020joint} & Model-based & GPI & DoA estimation & No & No & No & Both & Monostatic & 200 \\
\cite{li2013joint} & Model-based & GPI & DoA estimation & No & No & No & Both & Bistatic & 50--150 \\
\cite{li2015transmit} & Model-based & GPI & DoA estimation & No & No & No & Both & Bistatic & 60 \\
\cite{guo2017angle} & Model-based & GPI, ADI, MC & DoA estimation & No & No & No & Both & Bistatic & 256 \\
\cite{temiz2025unsupervised} & DL & N/A\makebox[0pt][l]{$^{\dagger}$} & Precoder design & N/A\makebox[0pt][l]{$^{\ddagger}$} & Yes & Yes & TX & Monostatic &  N/A\makebox[0pt][l]{$^{\ddagger}$} \\
\cite{chatelier2025physically} & MB-ML & GPI and ADI & DoA estimation & No & No & No & RX & N/A\makebox[0pt][l]{\Psuper} & 100 \\
\cite{konstantino2026unsupervised} & MB-ML   & ADI and NSV & Tracking DoA & Yes & No & No & RX & N/A\makebox[0pt][l]{\Psuper} & 20 / 200 \\
\cite{mateos2025unsupervised}\makebox[0pt][l]{$^{*}$} & MB-ML   & GPI & Single-target position estimation & No & Yes & Yes & RX & Monostatic &  1 \\ \midrule
This work & MB-ML & GPI and ADI & \begin{tabular}[c]{@{}c@{}}Multi-target position estimation \\ and precoder design\end{tabular} & Yes & Yes & Yes & Both & Monostatic &  1 \\ \bottomrule%
\end{tabular}%
    
\begin{tablenotes}
\footnotesize
\item $^{*}$Conference version of this paper.
\item $^{\dagger}$Not applicable: the impairments are modeled as random noise added to the estimated \ac{CSI}.
\item $^{\ddagger}$Not applicable: RX design is not considered.
\item {\Psuper}Not applicable: only the RX signal model is considered.
\end{tablenotes}
    
\end{threeparttable}
\end{table*}

\subsection*{Main Contributions} \label{subsec:main_contributions}
In this paper, we address the problem of calibrating the antenna arrays of a \ac{BS} that simultaneously performs monostatic sensing and communicates with a \ac{UE} using \ac{OFDM} signals. We consider an \ac{ISAC} \ac{BS} equipped with two \acp{ULA} used for transmission and reception, respectively. The \acp{ULA} are affected by \acp{GPI} and \acp{ADI}. We consider several targets in the \ac{FOV} of the \ac{BS} to sense and a communication \ac{UE} surrounded by scatterers. 

Our main contributions are summarized as follows:
\begin{itemize}
    \item \textbf{Unsupervised \ac{E2E} \ac{MB-ML} calibration:} 
    We propose for the first time an effective unsupervised calibration approach that simultaneously accounts for \ac{TX} and \ac{RX} impairments with coherent signals under a single snapshot.
    We calibrate the \acp{GPI} and \acp{ADI} while the ISAC \ac{BS} estimates the positions of the targets and communicates with the \ac{UE}. 
    Calibration is performed by parameterizing the steering vectors of the \acp{ULA} and optimizing them based on the sensing and communication loss functions computed from the received signals.
     As unsupervised sensing loss functions, we propose and compare: $(i)$ the negative maximum
    value of the \ac{ADM} of the received signal, i.e., the negative of the energy
    backscattered by the dominant scatterer in the scene, as each entry of the \ac{ADM}
    measures the energy that the \ac{RX} \ac{ULA} collects from one angle-range cell; and $(ii)$ the norm of the residual signal left once the dominant backscattered reflections have been subtracted, i.e., the portion of the received energy that the estimated scatterers do not
    account for.
    As an unsupervised communication loss function, we propose the negative estimated energy of the
    signal at the \ac{UE}, i.e., the negative of the signal energy that the transmit beam
    delivers to the UE across all subcarriers, which is set by the beamforming gain that the
    \ac{TX} \ac{ULA} attains in the direction of the UE.
    Compared to model-based calibration~\cite{liu2013unified, wang2019doa, wang2024robust}, our proposed approach works under coherent signals and \ac{OFDM}, which allows us to distinguish targets at the same angle but different ranges and to exploit the range-dependent phase difference across subcarriers to perform calibration. We consider simultaneous calibration of \acp{GPI} and \acp{ADI}.
    In contrast to~\cite{temiz2025unsupervised, chatelier2025physically}, the proposed method jointly compensates for impairments at the \ac{TX} and \ac{RX}, which we refer to as \ac{E2E} learning. Moreover, our steering vector parameterization reduces the number of learnable parameters compared to~\cite{temiz2025unsupervised}, and our results narrow the gap between \ac{SL} and \ac{UL} compared to~\cite{chatelier2025physically}.
    \item \textbf{Gradient-free channel backpropagation in ISAC:} 
    To perform \ac{E2E} learning, we consider the wireless channel as a function mapping the sensing beamformer and the communication symbols (input) to the received sensing and communication signals (output).
    Our proposed method approximates the gradient of the loss function with respect to the instantaneous channel function to avoid backpropagation of the gradient of the loss function through the channel function, 
    compared to other \ac{E2E} learning methods for calibration~\cite{mateos2025model}. 
    In a realistic scenario, the gradient of the channel function output with respect to its input is unknown.
    Moreover, the output of the channel function depends on the \ac{TX} impairments, which are specific to the \ac{ISAC} \ac{BS}, requiring that the steering vectors are dynamically updated based on new data.  
    Although gradient-free channel backpropagation has been applied to communications~\cite{aoudia19model}, we consider it for the first time in ISAC.
\end{itemize}

\subsubsection*{Organization}
The paper is organized as follows. In Sec.~\ref{sec:system_model}, we introduce the system model and the sensing and communication channels, including the model of the \acp{GPI} and \acp{ADI}. Sec.~\ref{sec:proposed_method} describes the proposed calibration method. In Sec.~\ref{sec:results}, we present the calibration results of the proposed approach and a comprehensive comparison with other approaches. Sec.~\ref{sec:conclusions} presents the main conclusions of this work and the outlook.

\subsubsection*{Notation}
Column vectors and matrices are denoted as boldface lowercase and uppercase letters, respectively. The transpose, conjugate and conjugate transpose of a matrix $\bm{A}$ are denoted as $\bm{A}^{\top}$, $\bm{A}^{*}$ and $\bm{A}^{\hermit}$, respectively. The $i$-th element of a vector is denoted as $[\bm{a}]_i$. The $i$-th column of a matrix is denoted as $[\bm{A}]_{:,i}$. The $\ell^2$-norm of a vector and the Frobenius norm of a matrix are denoted as $\norm{\bm{a}}$ and $\normfro{\bm{A}}$, respectively. The all-ones vector is denoted as $\bm{1}$. Sets are enclosed with curly brackets and denoted with calligraphic uppercase letters. The cardinality of a set $\Pset$ is denoted as $\abs{\Pset}$. The uniform distribution on the interval $[a,b]$ is denoted as $\U[a,b]$ and the uniform distribution over the set of values $\{a,b,c\}$ is denoted as $\U\{a,b,c\}$. The real-valued Gaussian distribution of mean $\bm{\mu}$ and covariance $\bm{\Sigma}$ is denoted as $\normal(\bm{\mu}, \bm{\Sigma})$. The circularly-symmetric complex Gaussian distribution with zero mean and covariance $\bm{\Sigma}$ is denoted as $\cnormal( \bm{\Sigma})$. The exponential distribution with mean $\mu$ is denoted as $\Exp(1/\mu)$. The expectation over a random variable $X$ is denoted as $\Expectation_X[\cdot]$.

\section{System Model} \label{sec:system_model}
We consider an ISAC \ac{BS} equipped with two \acp{ULA} of $K$ antenna elements in the same hardware platform that share the same clock and are used to transmit and receive signals. The \ac{ISAC} \ac{BS} transmits \ac{OFDM} signals with $S$ subcarriers to sense targets in the environment and to communicate with a \ac{UE}. It also receives the signal backscattered from the targets. In the following, we describe the individual sensing and communication received signal models, the joint ISAC model, and the effect of hardware impairments.
The notation of the most relevant terms in this section is summarized in Table~\ref{tab:sim_parameters}.

\subsection{Received Sensing Signal}
We consider at most $\Tmax$ targets and $\Cmax$ clutter scatterers in the scene at each transmission.
The signal backscattered from the targets and clutter that impinges on the \ac{RX} ULA
without hardware impairments is given by
\cite{5G_NR_JRC_analysis_JSAC_2022, MIMO_OFDM_ICI_JSTSP_2021,liu2026clutter}
\begin{align} \label{eq:sensing_model}
    \Ys = &\sum_{t=1}^T \alpha_t \bmarx(\theta_t) \bmatx^{\top}(\theta_t) \bmf [\bmx \odot \bmrho(\tau_t)]^{\top} \nonumber \\ 
    + &\sum_{c=1}^C \beta_c \bmarx(\theta_c) \bmatx^{\top}(\theta_c) \bmf [\bmx \odot \bmrho(\tau_c)]^{\top} +\bmW,
\end{align}
where $\Ys\in\complexset[K][S]$ collects the observations in the spatial-frequency domains, $T\in\{0, \ldots, \Tmax\}$ is the instantaneous number of targets in the scene, $C\in\{0,\ldots,\Cmax\}$ is the instantaneous number of clutter scatterers, $\alpha_t$ is the target complex channel gain, which depends on the distance to the target and the \ac{RCS} of the target, and $\beta_c$ is the clutter complex channel gain. The magnitude of $\alpha_t$ is given by the radar equation
\begin{align} \label{eq:mag_gain}
    |\alpha_t|^2 = \frac{\sigmarcst\lambda^2}{(4\pi)^3 \Rt^4},
\end{align}
while the phase is in the range $[0,2\pi)$. In \eqref{eq:mag_gain}, $\sigmarcst>0$ is the RCS of the $t$-th target, 
$\lambda$ is the carrier wavelength, and $\Rt$ is the distance between the ISAC \ac{BS} and the $t$-th target. 
The clutter complex channel gain follows the model of~\cite{liu2026clutter} and is distributed as $\beta_c\sim\cnormal(\sigmac^2)$, where $\sigmac^2$ determines the clutter
power.
The antenna and frequency-domain steering vectors $\bmax(\theta_t)$ and $\bmrho(\tau_t)$ in \eqref{eq:sensing_model} are defined as 
\begin{align}
    \bmax(\theta_t) &= [e^{\jmath2\pi\frac{K-1}{2\lambda}d\sin(\theta_t)},\ldots,e^{-\jmath2\pi\frac{K-1}{2\lambda}d\sin(\theta_t)}]^{\top}, \\
    \bmrho(\tau_t) &= [1, \ldots, e^{-\jmath2\pi (S-1) \Deltaf\tau_t}]^{\top},
\end{align}
where the subindex x denotes either \ac{TX} or \ac{RX}, $d=\lambda/2$ is the spacing between antenna elements, $\Deltaf$ is the subcarrier spacing in the OFDM signal, and $\tau_t=2\Rt/c$ and $\tau_c=2R_c/c$ are the round-trip delays to the $t$-th target and the $c$-th clutter scatterer, respectively. In \eqref{eq:sensing_model}, $\bmf$ denotes the ISAC precoder that steers the antenna energy into a particular direction, with power $\norm{\bmf}^2=P$, $\bmx$ are the communication symbols to be transmitted, drawn from a set $\Xset$ and satisfying  $\norm{\bmx}^2=S$, and $\bmW$ is the \ac{RX} \ac{AWGN}, with $[\bmW]_{i,j}\sim\cnormal(N_0S\Deltaf)$ and $N_0$ the noise power spectral density.
The angles and ranges of targets and clutter scatterers are within an uncertainty region, i.e.,  $\theta_t,\theta_c\in[\thetamin, \thetamax]$ and $\Rt,R_c\in[\Rmin, \Rmax]$. We assume that the ISAC \ac{BS} has knowledge of that uncertainty region.\footnote{\label{note:prior}Only a coarse knowledge of the uncertainty region is needed. It can be obtained from an initial beam sweeping stage, which is based on received-power measurements and is therefore robust to the array impairments described in Sec.~\ref{subsec:hwi}, i.e., the prior can be acquired before the array is calibrated.} We define the maximum achievable\footnote{Given that the actual sensing SCNR depends on the algorithm to compute the precoder $\bmf$ and the impairments, we upper-bound the sensing SCNR using the fact that $|\bmatx^{\top}(\theta)\bmf|^2\leq PK$.} sensing \ac{SCNR} as 
\begin{equation} \label{eq:scnr_sens}
\SCNRs = \Expectation_{\sigmarcst, R_t}[|\alpha_t|^2]/(N_0S\Deltaf/PK + \sigmac^2).    
\end{equation}

\subsection{Received Communication Signal}
We consider a single-antenna \ac{UE} that receives the signal emitted by the ISAC \ac{BS}. Between the \ac{BS} and the UE there are objects that scatter the signal in different directions. The signal impinging on the UE under no hardware impairments is given by~\cite{5G_NR_JRC_analysis_JSAC_2022, MIMO_OFDM_ICI_JSTSP_2021, Zohair_5G_errorBounds_TWC_2018}
\begin{align} \label{eq:comm_model}
    \yc = \sum_{t=1}^{\Ttilde} \alphatilde_t \bmatx^{\top}(\thetatilde_t) \bmf [\bmx \odot \bmrho(\tautilde_t)] + \bmw,
\end{align}
where $\yc\in\complexset[S]$ collects the observations in the frequency domain, $\Ttilde\in\{1,\ldots,\Ttildemax\}$ is the instantaneous number of paths; $\Ttildemax$ is the assumed maximum number of \ac{BS}--\ac{UE} paths; $\alphatilde_t, \thetatilde_t$, and $\tautilde_t$ are the complex channel gain, \ac{DoD}, and delay of the $t$-th path, respectively; and $\bmw\sim\cnormal(N_0S\Deltaf \bm{I}_S)$ is the \ac{RX} \ac{AWGN} at the UE. In \eqref{eq:comm_model}, $t=1$ represents the \ac{LoS} path between the \ac{BS} and the UE and $t>1$ are the \ac{BS}--scatterer--UE paths. The magnitude of the complex channel gain is modeled according to \cite[Eq. (45)]{Zohair_5G_errorBounds_TWC_2018} as
\begin{align}
    |\alphatilde_t|^2 = 
    \begin{cases}
	\lambda^2/ (4 \pi \Rtilde_1)^2 ,&~ t = 1   \\
	\lambda^2\sigmarcstt / [(4 \pi)^3 \Rtilde^2_{t,1} \Rtilde^2_{t,2}] ,&~ t > 1,
    \end{cases}
\end{align}
where $\Rtilde_1$ is the \ac{BS}--UE distance, $\sigmarcstt$ is the RCS of the scatterer for the $t$-th path and $\Rtilde_{t,1}$ and $\Rtilde_{t,2}$ are the \ac{BS}--scatterer and scatterer--UE distances, respectively. The UE is assumed to be within an uncertainty region, i.e., $\thetatilde_t\in[\thetatmin, \thetatmax]$ and $\Rtilde_1\in[\Rtilde_{\min}, \Rtilde_{\max}]$. We consider that the ISAC \ac{BS} has knowledge of the uncertainty region of the UE, obtained as described in footnote~\ref{note:prior}. Additionally, based on pilot data, we assume that the UE estimates the \ac{CSI} given by
\begin{align} \label{eq:csi}
    \bmkappa = \sum_{t=1}^{\Ttilde}\alphatilde_t \bmatx^{\top}(\thetatilde_t) \bmf \bmrho(\tautilde_t).
\end{align}
We define the maximum achievable communication \ac{SNR} as 
\begin{align}
    \SNRc = P\cdot K \cdot \Expectation_{\sigmarcstt, \Rtilde_1, \Rtilde_{t,1},\Rtilde_{t,2}}[|\alphatilde_t|^2]/(N_0S\Deltaf).
\end{align}

\subsection{ISAC Model}
The sensing model in \eqref{eq:sensing_model} and the communication model in \eqref{eq:comm_model} use the same precoder $\bmf$. This precoder is the ISAC precoder, which balances the power transmitted in the direction of the targets and the direction of the UE. 
It is computed according to~\cite{zhang2019multibeam} as
\begin{align} \label{eq:isac_precoder}
    \bmf = \sqrt{P}\frac{\sqrt{\omegar}\bmfs + \sqrt{1-\omegar}\bmfc}{\norm{\sqrt{\omegar}\bmfs + \sqrt{1-\omegar}\bmfc}},
\end{align}
where $P$ is the \ac{TX} power, $\omegar\in[0,1]$ is a hyper-parameter that selects how much power is radiated in the direction of the targets and $\bmfs\in\complexset[K]$ and $\bmfc\in\complexset[K]$ are the unit-norm sensing and communication precoders that illuminate the angle uncertainty regions $[\thetamin, \thetamax]$ and $[\thetatmin, \thetatmax]$, respectively. By sweeping over $\omegar$, we can explore the ISAC trade-offs of the system.

\subsection{Hardware Impairments} \label{subsec:hwi}
We consider hardware impairments in the two ULAs of the ISAC \ac{BS}. Particularly, we consider that they are affected by GPIs and ADIs.\footnote{\label{note:sic} Self-interference is not explicitly modeled in this work. We
assume that the isolation and cancellation stages of the monostatic
front-end~\cite{everett2014passive, smida2024inband} suppress the residual self-interference
below the dynamic range of the \ac{RX}, so that it does not saturate. The remaining static
leakage is then confined to the range-resolution cell at zero delay and can be excluded from
the sensing processing, provided that no target lies within it.} This changes the definition of the antenna steering vectors as 
\begin{align} \label{eq:impaired_steer_vec}
    \bmax(\theta_t;\bmgammax, \bmpx) = [&\gamma_{\mathrm{x},1}e^{-\jmath2\pi\frac{p_{\mathrm{x},1}}{\lambda}\sin(\theta_t)},\ldots, \nonumber\\
    &\gamma_{\mathrm{x},K}e^{-\jmath2\pi\frac{p_{\mathrm{x},K}}{\lambda}\sin(\theta_t)}]^{\top},
\end{align}
where $\bmgammax = [\gamma_{\mathrm{x},1}, \ldots, \gamma_{\mathrm{x},K}]^{\top}\in\complexset[K]$ and $\bmpx=[p_{\mathrm{x},1}, \ldots, p_{\mathrm{x},K}]^{\top}\in\realset[K]$ denote the vector of GPIs and antenna element positions, respectively. 
To make the hardware impairment model physically consistent, we consider that $p_{\mathrm{x},1} < \cdots < p_{\mathrm{x},K}$ and $|\gamma_{\mathrm{x},k}|\leq 1\ \forall k=1,\ldots, K$, i.e., the positions of the antenna elements are ordered in space and hardware impairments do not increase the radiated power of any antenna element.
We denote the \ac{TX} and \ac{RX} impairments as $\bmXitx=[\bmgammatx, \bmptx]$ and $\bmXirx=[\bmgammarx, \bmprx]$, respectively, and both impairments as $\bmXi = [\bmXitx, \bmXirx]$.

\begin{figure}[tb]
    \centering 
     \includestandalone[width=0.47\textwidth]{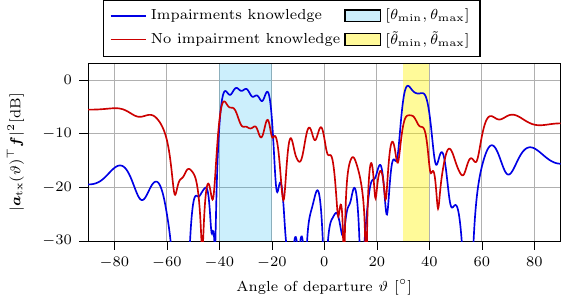}
    \caption{Precoder response $|\bmatx(\vartheta)^\top \bmf|^2$ under matched impairments and hardware impairments, for a sensing angular sector $[\thetamin, \thetamax] = [-40\degreee, -20\degreee]$ and a communication angular sector $[\thetatmin, \thetatmax] = [30\degreee, 40\degreee]$. The parameters $P$ and $\omegar$ in \eqref{eq:isac_precoder} are $P=0.1$ W, $\omegar=0.75$.}
    \label{fig:example_bf}
\end{figure}

\begin{example}[Effect of hardware impairments at the transmitter]
    Consider that the targets and the communication UE lie in the angular sectors $[\thetamin, \thetamax] = [-40\degreee, -20\degreee]$ and $[\thetatmin, \thetatmax]=[30\degreee, 40\degreee]$, respectively. In Fig.~\ref{fig:example_bf}, the precoder response $|\bmatx(\vartheta)^\top \bmf|^2$ is shown for $\vartheta\in[-90\degreee, 90\degreee]$ under matched impairments (the assumed and real steering vectors coincide) and hardware impairments (the assumed steering vector does not include impairments while the real steering vector does). In this example, we use one realization of the hardware impairment distributions given in Sec.~\ref{subsec:sim_param}. Details on how to compute $\bmfs$ and $\bmfc$ in \eqref{eq:isac_precoder} are given in Sec.~\ref{subsec:prop_tx}. Under hardware impairments, the energy of the precoder is diverted to undesired directions, while the matched precoder response focuses most of the energy at the desired angular sectors.
\end{example}

\begin{example}[Effect of hardware impairments at the receiver]
Consider that there are five targets in the environment. In Fig.~\ref{fig:omp_example}, we represent the \ac{ADM} of those targets together with their true positions (more details about the \ac{ADM} are discussed in Sec.~\ref{subsec:ul_sens_rx}). Particularly, Fig.~\ref{fig:omp_example_first} shows the \ac{ADM} when the signal is received and Fig.~\ref{fig:omp_example_last} shows the \ac{ADM} after all five targets have been estimated and  removed from the received signal following the \ac{OMP} algorithm. In Fig.~\ref{fig:omp_example_first}, we can observe that the maximum value of the \ac{ADM} is lower under hardware impairments compared to the matched case and the positions of the peaks are slightly displaced from the true positions. This effect caused the target with the highest range not to be removed and spurious peaks were falsely detected as targets. Moreover, Fig.~\ref{fig:omp_example_last} shows that the remaining \ac{ADM} after removing all targets has significantly lower values when impairments are matched compared to mismatched impairments.  This indicates that hardware impairments hinder target position estimation and it motivates the loss function of the proposed framework described in Sec.~\ref{subsec:ul_sens_rx}.

\begin{figure}[tb]
    \centering
    \begin{subfigure}{0.5\textwidth}
        \centering
        \includegraphics[width=\textwidth]{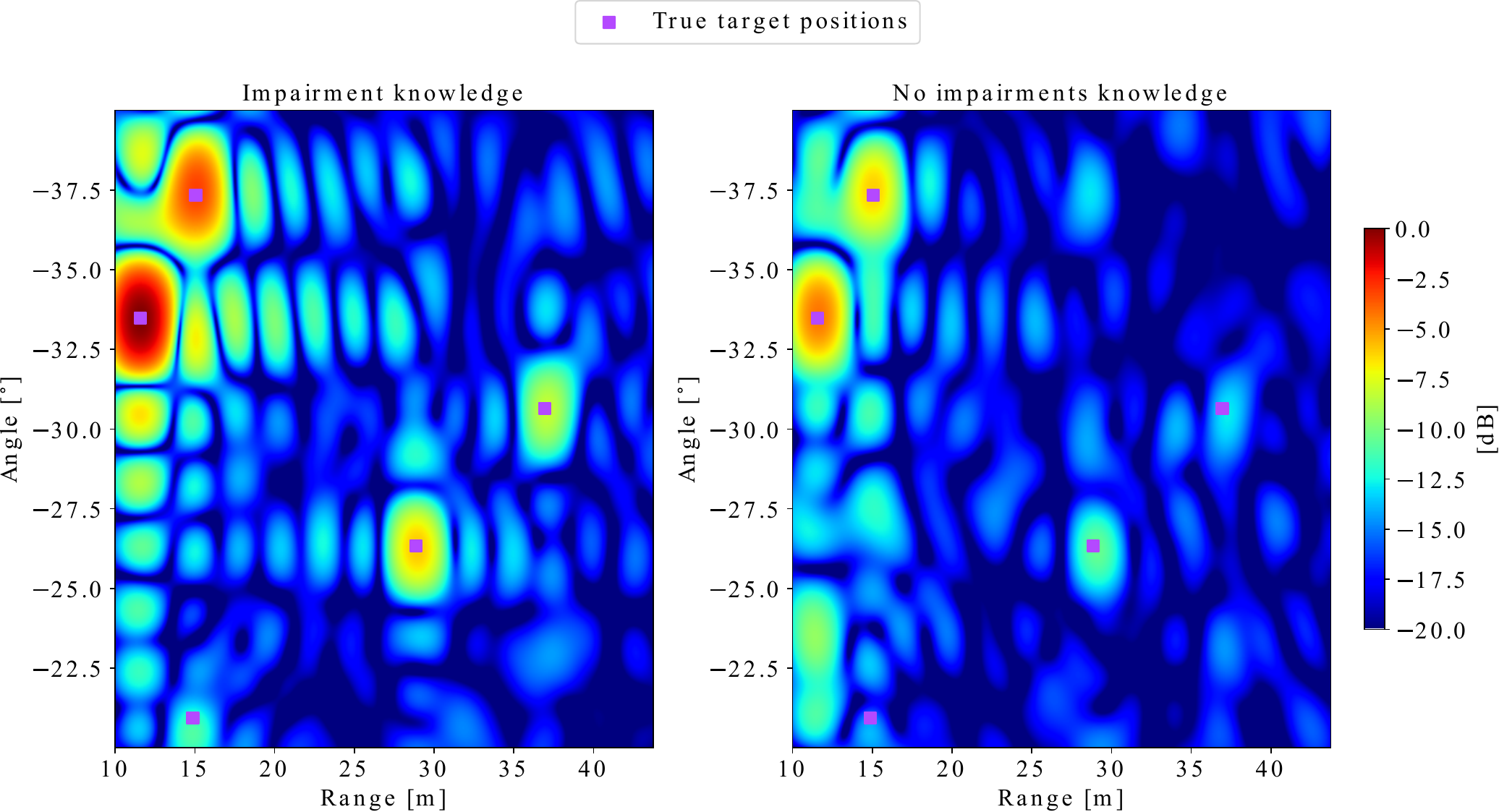}
        \caption{First OMP iteration.}
        \label{fig:omp_example_first}
    \end{subfigure}
    \begin{subfigure}{0.5\textwidth}
        \centering
        \includegraphics[width=\textwidth]{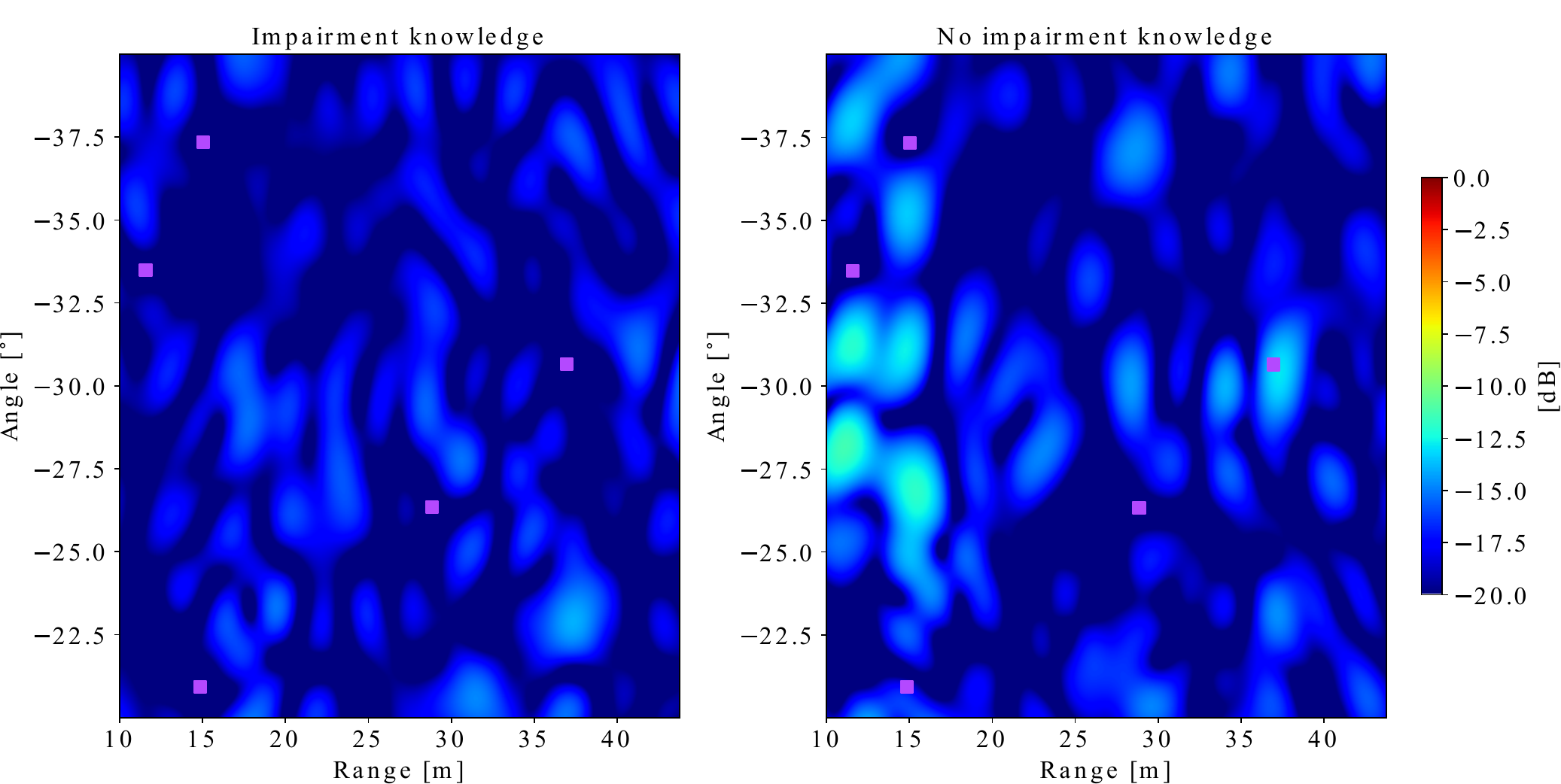}
        \caption{Residual after OMP iterations.}
        \label{fig:omp_example_last}
    \end{subfigure}
    \caption{\acp{ADM} for the first (top) and the last (bottom) iterations of the \ac{OMP} algorithm for five targets.}
    \label{fig:omp_example}
\end{figure}
    
\end{example}

\subsection{Problem Formulation}
Our objective is to enable the \ac{ISAC} system based on the above modeling  to compensate for the hardware impairments $\bmXi$. This implies that the system is expected to cope with the calibration errors while meeting the standard \ac{ISAC} objectives, i.e., accurately estimating: $(i)$ the number of targets $T$, $(ii)$ their positions $\{\theta_t, R_t\}_{t=1}^T$, and $(iii)$ the transmitted communication messages $\bmx$. The evaluation metrics to assess the performance of the ISAC system are described in Sec.~\ref{subsec:eval_metric}.

Specifically, at the \ac{TX}, prior information is available in the form of angular and range uncertainty regions for both the targets ($[\thetamin, \thetamax]$, $[\Rmin, \Rmax]$) and the \ac{UE} ($[\thetatilde_{\min}, \thetatilde_{\max}]$, $[\Rtilde_{\min}, \Rtilde_{\max}]$). These uncertainty regions can change for every new transmission and they determine the \ac{ISAC} precoder $\bmf$ in \eqref{eq:isac_precoder}. Moreover, the \ac{TX} has access to the transmission parameters, including the number of antennas $K$ and subcarriers $S$, the subcarrier spacing $\Deltaf$, the transmitted power $P$, the carrier wavelength $\lambda$, the \ac{ISAC} trade-off $\omegar$ in \eqref{eq:isac_precoder}, and the communication symbols $\bmx$. 

The sensing \ac{RX}, co-located with the \ac{TX} on the same hardware platform, receives the observations $\Ys$ in \eqref{eq:sensing_model}. Using these observations and the prior information available at the \ac{TX}, it estimates the number of targets and their positions. The \ac{UE} receives the observations $\yc$ in \eqref{eq:comm_model} and estimates the \ac{CSI} $\bmkappa$ in \eqref{eq:csi} based on pilot data. Using $\yc$ and $\bmkappa$, the \ac{UE} estimates the communication symbols.

\section{Proposed Method} \label{sec:proposed_method}
\begin{figure*}[t]
    \centering
    \includestandalone[width=\textwidth]{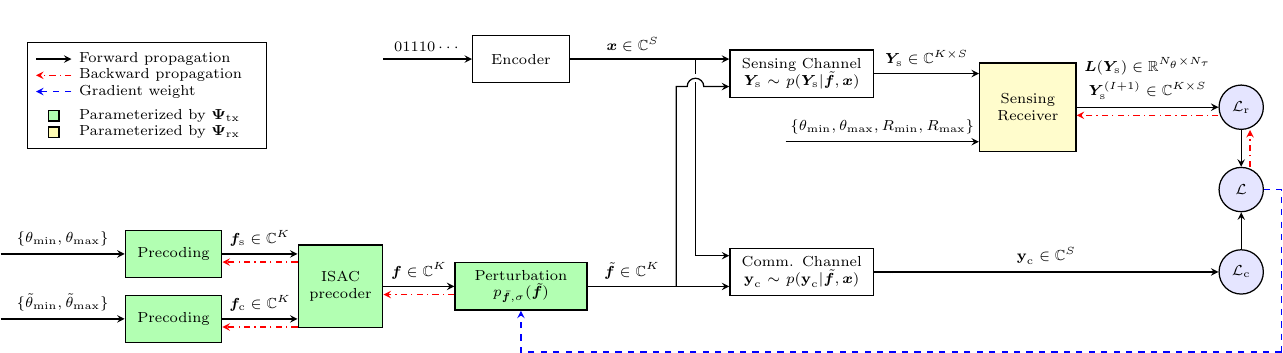}
    \caption{Calibration block diagram of the proposed approach. The gradient of the channel function is unknown and backward propagation is only possible through the transmitter and the sensing receiver.}
    \label{fig:training_block}
\end{figure*}

In this section, we introduce our method for unsupervised \ac{ISAC} array calibration.
The proposed calibration method is rooted in algorithmic steps that are parameterized and endowed with more degrees of freedom. Optimizing these parameters can account for impairments and improve overall \ac{ISAC} performance. 
Here, we describe the core algorithms and the corresponding parameterization at the \ac{TX}, sensing \ac{RX}, and communication \ac{RX}. At the \acp{RX}, we also detail the objective functions to minimize. 
To that aim, we first describe the individual \ac{TX} and \ac{RX} operations and how we process and combine the received signal to calibrate the \acp{ULA}. We finish the section with a description of the approach to avoid channel backpropagation. We represent in Fig.~\ref{fig:training_block} a block diagram of the calibration procedure.

\subsection{Transmitter} \label{subsec:prop_tx}
The goal of the \ac{TX} is to compute a precoder $\bmf$ that illuminates the sensing and communication angular uncertainty regions according to \eqref{eq:isac_precoder}. 
Here, we describe how to compute the individual $\bmfs$ and $\bmfc$ following similar operations, which we later combine according to \eqref{eq:isac_precoder}. We here generally denote $\bmfs$ or $\bmfc$ as $\bmfx$ and $[\thetabmin, \thetabmax]$ as the uncertainty angular sector for either sensing or communications.
We design the precoder $\bmfx$ in \eqref{eq:sensing_model} and \eqref{eq:comm_model} by noting that the solution that maximizes the transmitted energy to a particular angle $\theta$, i.e.,  
\begin{align}
    \arg\max_{\bmfx} |\bmatx^{\top}(\theta) \bmfx|&, \\
    \mathrm{s.t.}\  \norm{\bmfx}^2=1&, \notag
\end{align}
is given by $\bmfx=\bmatx^*(\theta)/\norm{\bmatx(\theta)}$. This solution implies knowledge of the target angle $\theta$, which is not available. Since we only have knowledge of the angular sector $[\thetabmin, \thetabmax]$, we consider an angular grid $\{\bartheta_i\}_{i=1}^{\Ntheta}$ that covers the field-of-view of the \ac{ISAC} \ac{BS} $[-\thetafov,\thetafov]$ and compute the precoder\footnote{The precoder in \eqref{eq:tx_proposed} is generally not optimal for a random $[\thetabmin, \thetabmax]$. We follow \eqref{eq:tx_proposed} for its simplicity and exploration of more optimal precoding algorithms for the proposed scenario is left as future work.} following~\cite{sun2026adaptive} as
\begin{align} \label{eq:tx_proposed}
    \bmfx(\bmPsitx) = \frac{\sum_{\bartheta_i\in[\thetabmin, \thetabmax]} \bmatx^*(\bartheta_i;\bmPsitx)}{\norm{\sum_{\bartheta_i\in[\thetabmin, \thetabmax]} \bmatx^*(\bartheta_i;\bmPsitx)}},
\end{align}
where $\bmfx$ is parameterized by $\bmPsitx$. The parameterized steering vector is expressed as
\begin{align} \label{eq:steer_vector_param}
    \bmatx(\bartheta;\bmPsitx) = [&\xi_{\mathrm{tx},1}e^{-\jmath2\pi\frac{\omega_{\mathrm{tx},1}}{\lambda}\sin(\bartheta)},\ldots, \nonumber\\
    &\xi_{\mathrm{tx},K}e^{-\jmath2\pi\frac{\omega_{\mathrm{tx},K}}{\lambda}\sin(\bartheta)}]^{\top},
\end{align}
where $\bmxitx = [\xi_{\mathrm{tx},1}, \ldots, \xi_{\mathrm{tx},K}]^{\top}$ and $\bmomegatx = [\omega_{\mathrm{tx},1}, \ldots, \omega_{\mathrm{tx},K}]^{\top}$ are learnable parameters and $\bmPsitx=[\bmxitx, \bmomegatx]$. We constrain the parameters such that $|\xi_{\mathrm{tx},k}|\leq 1,\ \forall k=1,\ldots,K$ and $\omega_{\mathrm{tx},1} < \cdots < \omega_{\mathrm{tx},K}$. We distinguish between the learnable parameters $\bmPsitx$ that can change to calibrate the \ac{ULA} and the actual impairments $\bmXitx$ that are fixed and inherent to the \ac{ULA}.

\subsection{Sensing Receiver} \label{subsec:ul_sens_rx}
To detect multiple targets and estimate their positions, we formulate the multi-target sensing problem as a sparse signal recovery problem and leverage the \ac{OMP} algorithm \cite{Mallat1993, OMP_TIT_2007,OMP_mmWave_2016} to solve it. We discretize the angular and delay uncertainty regions $[\thetamin, \thetamax], [\taumin, \taumax]$ and construct the angular and delay-domain dictionaries as
\begin{align}
    \bmPhia &= [\bmarx(\theta_1;\bmPsirx), \ldots, \bmarx(\theta_{\Ntheta};\bmPsirx)] \in \complexset[K][\Ntheta], \label{eq:angle_dict} \\
    \bmPhid &= \bmx\bm{1}^{\top} \odot [\bmrho(\tau_1), \ldots, \bmrho(\tau_{\Ntau})]\in\complexset[S][\Ntau], \label{eq:delay_dict}
\end{align}
where $\bmPsirx = [\bmxirx, \bmomegarx]$ follows analogous definitions and constraints as $\bmPsitx$ in~\eqref{eq:steer_vector_param}.
Note that since we assume a co-located ISAC \ac{BS}, the transmitted communication symbols are known during reception. 
The frequency dimension is instrumental to the proposed calibration. The delay dictionary in~\eqref{eq:delay_dict} resolves reflections that share the same angle but differ in range, so that a mismatch in the \ac{ADM} can be ascribed to the array impairments rather than to unresolved reflections.
Using \eqref{eq:angle_dict} and \eqref{eq:delay_dict}, we can express the received observations $\Ys$ in \eqref{eq:sensing_model} as
\begin{align}
    \Ys = \sum_{i=1}^{\Ntheta} \sum_{j=1}^{\Ntau} [\bmS]_{i,j} [\bmPhia]_{:,i} ([\bmPhid]_{:,j})^{\top} + \bmW,
\end{align}
where $\bmS\in\complexset[\Ntheta][\Ntau]$. The goal is to estimate the $T$-sparse matrix $\bmS$ under the assumption $T \ll \Ntheta\Ntau$. The OMP algorithm is summarized in Algorithm~\ref{alg:omp}. 
\begin{algorithm}[!tb]
\caption{{OMP for Multi-Target Sensing}}
\label{alg:omp}
\begin{algorithmic}[1]
    \State \textbf{Input:} Observation $\Ys$ in \eqref{eq:sensing_model}, angular grid $\{\theta_i\}_{i=1}^{\Ntheta}$, delay grid $\{\tau_j\}_{j=1}^{\Ntau}$, and termination threshold $\delta$. 
    \State \textbf{Output:} Set {$\Psethat$}, which contains the angle and delay estimates of multiple targets $\{ (\thetahat_{t}, \tauhat_{t})\}_{t=1}^{I}$.
 \State \textbf{Initialization:} Set $I=0$, ${\Psethat} = \varnothing$, $\bmAa = \bmAd = [ ~ ]$. \\
 Set the residual to $\Ys^{(0)} = \Ys$.\\
 Compute dictionaries $\bmPhia$ and $\bmPhid$ according to \eqref{eq:angle_dict} and \eqref{eq:delay_dict}, respectively. \\
 Compute the \ac{ADM} $\bmL(\Ys^{(I)}) = |\bmPhia^{\hermit} \Ys^{(I)} \bmPhid^\ast|^2$.
 \State \textbf{while} $\max_{i,j} [\bmL(\Ys^{(I)})]_{i,j} > \delta$ 
    \Indent
    \State \label{line_argmax} Angle-delay detection: 
     \begin{align} \label{eq:omp_argmax}
         (\ihat, \jhat) = \arg \max_{i,j} [\bmL(\Ys^{(I)})]_{i,j} ~.
     \end{align}
    \State $(\thetahat_I, \tauhat_I) \gets (\theta_{\ihat}, \tau_{\jhat})$.
    \State \label{line_update_pair} Update angle-delay pairs: ${\Psethat} \gets {\Psethat} \cup \{ (\thetahat_I, \tauhat_I ) \}$.
    
    \State \label{line_update_set} Update atom sets: 
    \begin{align}
        \bmAa &\gets [ \bmAa ~  [\bmPhia]_{:, \ihat} ] ~,  \\
        \bmAd &\gets [ \bmAd ~  [\bmPhid]_{:, \jhat} ] ~.
    \end{align}
\State \label{line_update_gain} Update gain estimates: 
\begin{align}
 \bmalphahat = \arg \min_{\alpha_1,\ldots,\alpha_{I+1}} \norm{ \Ys - \sum_{t=1}^{I+1} \alpha_t [\bmAa]_{:, t} ([\bmAd]_{:, t})^\top  }_F^2 ~.
\end{align}
    \State Update residual: 
    \begin{align} \label{eq:update_residual}
        \Ys^{(I+1)} = \Ys - \sum_{t=1}^{I+1} \hat{\alpha}_t [\bmAa]_{:, t} ([\bmAd]_{:, t})^\top ~.
    \end{align}
    \State \label{line_update_I} $I = I + 1$.
 \EndIndent
 \State \textbf{end while}
\end{algorithmic}
\normalsize
\end{algorithm}
Based on the OMP algorithm, we propose two \ac{UL} loss functions that require no labeled data (in the form of the true number of targets $T$, their angles $\theta_t$, and delays $\tau_t$) to optimize $\bmPsi = [\bmPsitx, \bmPsirx]$.
\subsubsection{Maximize the \ac{ADM} Response}
The \ac{ADM} $\bmL(\Ys)$ contains high values (peaks) at the true target locations under no hardware impairments. However, the effect of the impairments shifts the peaks and decreases the magnitude of the \ac{ADM}, as observed in Fig.~\ref{fig:omp_example}. We then propose to maximize the maximum response of the \ac{ADM}, expressed in terms of a loss function as
\begin{align} \label{eq:loss_r_1}
    \Lr(\bmPsi) = -\max_{i,j}[\bmL(\Ys(\bmPsi))]_{i,j},
\end{align}
where we explicitly included the dependency of $\Ys$ on $\bmPsi$ ($\bmPsitx$ is embedded in $\bmS$ from the precoder $\bmfx$ in \eqref{eq:tx_proposed} and $\bmPsirx$ is included in $\bmPhia$ in \eqref{eq:angle_dict}). This loss function was first proposed in~\cite{mateos2025unsupervised} for a simpler \ac{ISAC} scenario. The loss in \eqref{eq:loss_r_1} does not require estimating the targets using the \ac{OMP} algorithm during training, only during inference once the \ac{ISAC} \ac{BS} has been calibrated.

\subsubsection{Minimize the OMP Residual Norm}
According to \eqref{eq:update_residual} in the OMP algorithm, the residual in the last iteration should not contain contributions from any of the targets and only noise should remain. We propose to minimize the norm of the residual in \eqref{eq:update_residual}
\begin{align} \label{eq:loss_r_2}
    \Lr(\bmPsi) = \norm{\Ys^{(I+1)}(\bmPsi)}_F^2.
\end{align}
The number of iterations $I$ will be discussed in Sec.~\ref{sec:results}. This loss function requires estimating the positions of the targets, increasing the computational complexity compared to the loss in~\eqref{eq:loss_r_1}.

\subsection{Communication Receiver} \label{subsec:ul_comm_rx}
According to the signal model in \eqref{eq:comm_model} and the CSI estimated by the UE in \eqref{eq:csi}, the communication signal received by the UE can be equivalently expressed as $\yc = \bmkappa\odot\bmx + \bmw$. Assuming that the symbols $\bmx$ are drawn from an equiprobable distribution, the optimal decoder corresponds to the subcarrier-wise maximum likelihood estimator
\begin{equation} \label{eq:ml_detection}
    [\hat{\bmx}]_s = \arg\min_{x\in\Xset} |[\yc]_s - [\bmkappa]_sx|^2.
\end{equation}

To propose a \ac{UL} loss to calibrate the \ac{TX} \ac{ULA}, we note that the impairments affect the TX ULA, which affect how the TX energy is steered in the direction of the UE and decrease the SNR received by the UE. We then propose to maximize the energy of the received signal by the UE, or in terms of a loss function
\begin{align} \label{eq:loss_comm}
    \Lc(\bmPsitx) = -\norm{\yc(\bmPsitx)}^2.
\end{align}

\subsection{ISAC Calibration as \ac{UL}}
\label{subsec:isac_cal_as_ul}
In Secs.~\ref{subsec:ul_sens_rx} and \ref{subsec:ul_comm_rx}, we described the individual sensing and communication \ac{UL} functions. Based on these formulations, we can cast the overall system as an \ac{MB-ML} model whose parameters are $\bmPsi$. Accordingly, the aforementioned loss functions allow calibrating the \ac{ISAC} systems as a form of \ac{UL}.

To optimize a joint ISAC objective, we consider a feasible impairment set 
\begin{align} \label{eq:imp_set}
    \mathcal{I}=\{\bmPsi~:~\omegax[1] < \cdots < \omegax[K], \abs{\xix[k]} \leq 1, \forall k\in\{1,\ldots, K\}\}
\end{align}
that enforces the parameters to  the physical constraints of the hardware impairments in \eqref{eq:impaired_steer_vec}. In~\eqref{eq:imp_set}, $\xix, \omegax$ refer to either $\xitx, \omegatx$ in \eqref{eq:steer_vector_param} or $\xirx, \omegarx$ in \eqref{eq:angle_dict}. Considering that the angular uncertainty sectors $\thetaint=\{[\thetamin, \thetamax], [\thetatmin, \thetatmax]\}$ and the communication symbols $\bmx$ are randomly distributed and given by higher-layer protocols, we formulate the joint optimization problem as
\begin{align} 
    \arg\min_{\bmPsi}\ & \Lcal(\bmPsi), \label{eq:isac_opt_problem} \\
    \mathrm{s.t.}\ &\bmPsi\in\mathcal{I}, \label{eq:isac_opt_problem_constraint}
\end{align}
where $\Lcal(\bmPsi)= \Expectation_{\bmzeta, \Ys, \yc}[ \etar \Lr(\bmPsi) + (1-\etar)\Lc(\bmPsitx)]$, $\bmzeta=\{\thetaint, \bmx\}$, and
$\etar$ is a hyper-parameter that balances the sensing and communication losses.

Given that the \ac{ISAC} \ac{BS} is continuously operating while calibration takes place, we tackle problem \eqref{eq:isac_opt_problem} via \ac{POGD}~\cite{wood2022online} as follows: $(i)$ we initialize the optimization with a parameter estimate $\bmPsi^{(0)}$; $(ii)$ in the $i$-th iteration of \ac{POGD}, we consider a new random data set $\mathcal{B}_i=\{\bmzeta_j, \Ys[j], \yc[j]\}_{j=1}^B$ from $B$ independent transmissions\footnote{\label{note:snapshots}The $B$ transmissions in $\mathcal{B}_i$ constitute a minibatch of the stochastic optimization and not a processing interval: each loss sample in~\eqref{eq:isac_loss_approx} is evaluated from its own single snapshot, and no statistic is accumulated across the $B$ transmissions. They are processed in parallel for computational efficiency and averaged only to reduce the variance of the gradient estimate. The proposed algorithm can operate with $B=1$ at the cost of a noisier gradient.} and approximate the \ac{ISAC} loss function as
\begin{align} \label{eq:isac_loss_approx}
    \Lcal(\bmPsi) \approx \Lcal_{\mathcal{B}_i}(\bmPsi) = \frac{1}{B}\sum_{j=1}^B &\etar \Lr(\bmPsi;\Ys[j]) \nonumber \\
    &+ (1-\etar)\Lc(\bmPsitx;\yc[j]);
\end{align}
$(iii)$ we update the parameters $\bmPsi^{(i)}$ based on the gradient $\nabla_{\bmPsi}\Lcal_{\mathcal{B}_i}(\bmPsi)$; and $(iv)$ the updated parameters $\bmPsi^{(i)}$ are projected onto the feasible set $\mathcal{I}$, namely, $\{\omegax[k]\}_{k=1}^K$ are ordered and $\xix[k]$ are normalized if $|\xix[k]|>1$ for any $k$. 
Note that the optimization problem in~\eqref{eq:isac_opt_problem} does not guarantee that the parameters $\bmPsi$ converge to the true impairments $\bmXi$. The objective is to improve the \ac{ISAC} performance of the considered system in terms of the sensing and communication metrics defined in Sec.~\ref{subsec:eval_metric} and their attainable tradeoffs evaluated in Sec.~\ref{subsec:isac_results}.

\begin{remark} \label{remark:complexity}
    The computational complexity  of~\eqref{eq:isac_loss_approx} is dominated by the \ac{OMP} algorithm. According to~\cite{OMP_TIT_2007}, the complexity of OMP in our case is $\bigO(IS\Ntheta\Ntau)$, assuming that $\Ntau\gg K$ (as outlined in Sec.~\ref{subsec:sim_param}), where $I$ is the number of OMP iterations.\footnote{\label{note:omp_complexity} Further reduction of the OMP complexity can be performed following, for instance,~\cite{jost2006tree, shad2025newtonized}.}
\end{remark}

\subsection{Gradient-Free Channel Backpropagation} \label{subsec:gf_backprop}
The proposed loss functions in \eqref{eq:loss_r_1}, \eqref{eq:loss_r_2}, and \eqref{eq:loss_comm} are computed at the sensing or communication \acp{RX}, but they depend on $\bmPsitx$. 
Consider, for example, the communication loss $\Lc$ in \eqref{eq:loss_comm}. The received signal $\yc$ is a random variable following a \ac{PDF} $p(\yc|\bmf(\bmPsitx), \bmx)$ according to \eqref{eq:comm_model}.
Backpropagating to optimize $\bmPsitx$ would require knowing the gradient $\nabla_{\bmPsitx}p(\yc|\bmf(\bmPsitx), \bmx)$. 
However, in a real scenario, $p(\yc|\bmf(\bmPsitx), \bmx)$ is unknown and it may include non-differentiable elements such as quantization at \ac{TX} or \ac{RX}, making the computation of its gradient infeasible.\footnote{\label{note:csi_gradient}The \ac{CSI} in~\eqref{eq:csi}, which the UE already estimates for detection, does not circumvent this issue. It describes the channel as seen through the precoder, and therefore does not reveal how the loss would change if the precoder, and hence $\bmPsitx$, were modified. Recovering that dependence would require an explicit model of the propagation and of the transceiver chain, which would bias the resulting gradient whenever the model does not hold.}
To circumvent this issue, we adopt the model-free \ac{E2E} training approach of \cite{aoudia19model} for our system, which we describe in the following example for the case of optimizing $\bmPsitx$ based on the communication loss. 

Considering that the angular uncertainty sectors $\thetaint=\{[\thetamin, \thetamax], [\thetatmin, \thetatmax]\}$ and the communication symbols $\bmx$ are randomly distributed, the expected communication loss function to minimize is
\begin{align} \label{eq:exp_loss}
    \Lbar_{\mathrm{c}}(\bmPsitx) = \E_{\bmzeta} \bigg[ \int \Lc(\bmPsitx) p(\yc|\bmf(\bmPsitx),\bmx) \d \yc \bigg],
\end{align}
where $\bmzeta=\{\thetaint, \bmx\}$ and $\Lc(\bmPsitx)$ is the instantaneous loss in~\eqref{eq:loss_comm} for one realization of $\yc$. The gradient $\nabla_{\bmPsitx}\Lbar_{\mathrm{c}}(\bmPsitx)$ requires computing $\nabla_{\bmu}p(\yc|\bmu,\bmx)|_{\bmu=\bmf(\bmPsitx)}$, which is not available in practice. As a workaround, we consider that the precoder $\bmf(\bmPsitx)$ is perturbed and  distributed according to a random variable $\bmfpert(\bmPsitx)$ with a \ac{PDF} $p_{\bmfbar, \sigma}(\bmfpert(\bmPsitx))$, 
where $\bmfbar=\bmf(\bmPsitx)$ and $\sigma$ are the expected value and standard deviation of $\bmfpert(\bmPsitx)$, respectively. The details of the precoder perturbation are given in Sec.~\ref{subsec:sim_param}. Then, the loss in \eqref{eq:exp_loss} becomes
\begin{align}
    \Lbar_{\mathrm{c}}(\bmPsitx) = \E_{\bmzeta} \bigg[&\int p_{\bmfbar, \sigma}(\bmfpert(\bmPsitx)) \nonumber\\
    &\int \Lc(\bmPsitx) p(\yc|\bmfpert(\bmPsitx),\bmx) \d \yc \d\bmfpert \bigg],
\end{align}
and the gradient with respect to $\bmPsitx$
\begin{align}
    \nabla_{\bmPsitx}\Lbar_{\mathrm{c}}&(\bmPsitx) \nonumber \\
    = 
    \E_{\bmzeta} \bigg[&\int \nabla_{\bmPsitx} p_{\bmfbar, \sigma}(\bmfpert(\bmPsitx)) \nonumber\\
    &\int \Lc(\bmPsitx) p(\yc|\bmfpert(\bmPsitx),\bmx) \d \yc \d\bmfpert \bigg] \nonumber \\
    =\E_{\bmzeta} \bigg[&\int \nabla_{\bmPsitx} \log(p_{\bmfbar, \sigma}(\bmfpert(\bmPsitx))) \nonumber\\
    &\int \Lc(\bmPsitx) p_{\bmfbar, \sigma}(\bmfpert(\bmPsitx)) p(\yc|\bmfpert(\bmPsitx),\bmx) \d \yc \d\bmfpert \bigg] \label{eq:grad_free_log_trick}, 
\end{align}
where in~\eqref{eq:grad_free_log_trick} we used the log-trick $\nabla_{\bmu}g(\bmu) = g(\bmu) \nabla_{\bmu}\log(g(\bmu))$. Considering that $p_{\bmfbar, \sigma}(\bmfpert(\bmPsitx))  p(\yc|\bmfpert(\bmPsitx),\bmx) = p(\yc,\bmfpert(\bmPsitx)|\bmx)$, we have that 
\begin{align} \label{eq:grad_free}
    \nabla_{\bmPsitx}\Lbar_{\mathrm{c}}(\bmPsitx) = \E_{\thetaint, \bmfpert, \yc} \bigg[&\Lc(\bmPsitx) \nonumber \\
    &\nabla_{\bmPsitx} \log(p_{\bmfbar, \sigma}(\bmfpert(\bmPsitx))) \bigg| \bmx \bigg].
\end{align}
In \eqref{eq:grad_free}, one only needs knowledge of the gradient of the logarithm of the \ac{PDF} of the perturbed precoder, which is available at the \ac{TX} side. The loss function $\Lc(\bmPsitx)$ has the role of weighting the gradients in \eqref{eq:grad_free} to yield suitable parameters $\bmPsitx$ (represented as a blue arrow in Fig.~\ref{fig:training_block}).\footnote{\label{note:feedback}We consider an ideal link from the UE to the ISAC transceiver for transmitting the loss function, in order to isolate the effect of the TX impairments and the multipath communication channel in~\eqref{eq:comm_model}. A distorted feedback signal has been shown to yield performance similar to that of an ideal feedback signal in~\cite{aoudia19model, song2020learning}.} 
Compared with knowledge of the true gradient, the optimization of the \ac{TX} parameters using the
approximated gradient requires more training samples~\cite{aoudia19model}, since the variance of the
estimator in~\eqref{eq:grad_free} is reduced by averaging over independent precoder perturbations, each
of which requires a separate transmission. The \ac{TX} parameters hence determine the observation window
needed for calibration, whereas the \ac{RX} parameters are optimized from the sensing losses
in~\eqref{eq:loss_r_1} and~\eqref{eq:loss_r_2}, which are differentiable with respect to $\bmPsirx$ and can be evaluated repeatedly on the same received signal.
The form of the gradient in~\eqref{eq:grad_free} is equivalent to the policy gradient estimator of~\cite{williams1992simple}, which guarantees that the expected value (over transmissions) of the direction in which the parameters $\bmPsitx$ are updated coincides with the expected value of the true gradient of the loss function. Moreover, the approximate gradient in~\eqref{eq:grad_free} was proven to tend to the true gradient in~\cite[Appendix A]{aoudia19model}.
The precoder $\bmfpert(\bmPsitx)$ follows a random distribution only harnessed during training. 
The perturbation of the precoder is a training mechanism only. Given that the considered \acp{GPI} and \acp{ADI} are intrinsic to the hardware platform and do not change over time, once the calibration has converged we freeze the learned parameters $\bmPsi$, and the precoder, which follows~\eqref{eq:isac_precoder}, no longer undergoes any random perturbation. The frozen parameters are reused for every subsequent transmission: since they describe the \acp{ULA} and not the propagation environment, they do not depend on the angular uncertainty regions, on the number of targets, or on the channel realization.

\section{Experimental Study}   \label{sec:results}
\subsection{Simulation Parameters} \label{subsec:sim_param}
The main simulation parameters of the ISAC scenario are outlined in Table~\ref{tab:sim_parameters}.\footnote{\label{footnote:antennas}We consider a \ac{BS} with 64 antennas since their practical feasibility has been evaluated for the 5th generation of wireless systems~\cite{guo2019system} and more antennas are being considered for 6G~\cite{bjornson2025enabling}. This is also the regime in which calibration is most critical, as the penalty incurred by uncompensated impairments grows with the array size. The description of the proposed approach is general and can be adapted to any number of antennas.}
For the experimental study, we consider that the inter-antenna position impairments follow the model of \cite{shmuel2025subspacenet}, i.e., $\bmpx = \bmpideal + \bmvarepsilonp$, where $\bmpideal = [-(K-1)\lambda/4, \ldots, (K-1)\lambda/4]^{\top}$ corresponds to the positions of an ideal ULA with half-wavelength spacing centered around zero and $\bmvarepsilonp$ is a perturbation of the ideal positions. Additionally, the model of the GPIs is similar to \cite{jiang2013two}, but we consider that the magnitude of the impairments cannot be greater than 1, i.e., there are no amplification components when considering GPIs. To avoid the ambiguity between \acp{GPI} and target angle, we assume that the first antenna element in both TX and RX ULA has known gain, phase, and position (reference sensor). Accordingly, the feasible set in~\eqref{eq:isac_opt_problem_constraint} is replaced by
\begin{align}
    \Isettilde = \Iset \cap \{\bmPsi:\omegax[1] = p_{\mathrm{x},1}, \xix[1]=\gamma_{\mathrm{x},1}\}.
\end{align}

To compute the received communication signal in \eqref{eq:comm_model}, scatterers are distributed to ensure that there is a \ac{LoS} path between \ac{TX} and \ac{RX} and that the cyclic prefix $\Tcp$ is larger than the delay spread, i.e., $\Tcp\geq(\max_t\{\Rtilde_{t,1}+\Rtilde_{t,2}\}-\Rtilde_1)/c$, $\forall t>1$. The cyclic prefix is set to $7$ percent of the elementary symbol duration in accordance with 3GPP specifications \cite[Ch.~5.3.1]{TS_38211}. Regarding the sensing estimation of targets, the angular grid $\{\theta_i\}_{i=1}^{\Ntheta}$ in Algorithm~\ref{alg:omp} spans $[\thetamin, \thetamax]$ and the delay grid $\{\tau_j\}_{j=1}^{\Ntau}$ spans $[2\Rmin/c, 2\Rmax/c]$. 

For the optimization of $\bmPsi$, we initialize the learnable parameters (except for the first sensor) as $\bmPsi^{(0)}=[\bm{1},\bmpideal, \bm{1}, \bmpideal]$, which corresponds to the case of no impairment knowledge. Moreover, the \ac{ISAC} precoder in \eqref{eq:isac_precoder} is perturbed as $\bmfpert=\sqrt{1-\sigma^2}\bmf+\bmvarepsilonf$, followed by a normalization step so that $\norm{\bmfpert}^2=1$.\footnote{\label{note:normalization}For gradient estimation, we regard the normalization as part of the unknown channel and evaluate~\eqref{eq:grad_free} with the Gaussian density of the precoder before normalization, since its log-likelihood has a simple closed-form expression.} We leverage the Adam optimizer~\cite{kingma2015adam} where we also use a scheduler in our proposed approach with the default PyTorch hyper-parameters except for a decaying factor of 0.5, a patience of 1000 iterations, and a cool-down of 500 iterations.
We explored the values $\{10^{-2}, 10^{-3}, 10^{-4}\}$ for the learning rate, $\{1000, 5000, 10000, 20000\}$ training iterations, and $\{50, 500, 4000\}$ samples for the batch size. We outline in Table~\ref{tab:sim_parameters} the hyper-parameters that yield the best results with the least number of iterations during training.\footnote{\label{note:optimization} We do not decrease the value of $\sigma$ over iterations for simplicity given the results presented in Secs.~\ref{subsec:sensing_results}--\ref{results:generalization}. Moreover, assuming that each transmission corresponds to one OFDM symbol, the total calibration time is approximately $3$ minutes for the parameters in Table~\ref{tab:sim_parameters}. Since the impairments are static and calibration runs online without dedicated resources, this time is not critical. It could be further reduced, for instance, by reusing received signals for the optimization of the RX parameters, which is left for future work.}\footnote{\label{note:code} Source code to reproduce all numerical results in this paper is available at \url{https://github.com/josemateosramos/Unsupervised_array_calibration_ISAC}}

\begin{table}[t]
\centering%
\caption{Simulation parameters}%
\label{tab:sim_parameters}%
{%
\begin{tabular}{c|c|c}%
\toprule%
Symbol & Meaning & Value  \\
\midrule
\midrule
$S$ & Number of subcarriers  & 256 \\ 
$\lambda$ & Wavelength & $5$ mm  \\
$\Deltaf$ & Subcarrier spacing & $240$ kHz  \\
$\Tcp$ & Cyclic prefix time & $0.29\, \rm{\mu s}$  \\
$P$ & Transmitted power & $0.1$ W  \\ \midrule

$K$  & Antennas in the \acp{ULA} & 64 \\
$\thetafov$  & Angular field-of-view of \ac{TX} & $\pi/2$ \\
$[\bmvarepsilonp]_k$ & \Ac{ADI} perturbation & $\U[-\lambda/5, \lambda/5]$ \\ 
$|[\bmgammax]_k|$ & \multirow{2}{*}{\acp{GPI}}  & $\U[0.95,1]$  \\ 
$\measuredangle{([\bmgammax]_k})$ &  & $\U[-\pi/2,\pi/2]$ \\
$\bmvarepsilonf$ & \multirow{2}{*}{Precoder perturbation} & $\cnormal(\sigma^2\bm{I})$ \\
$\sigma$ &  & $0.05$ \\\midrule

$\Tmax$ & Maximum sensing targets & 5   \\
$\Ttildemax$ & Maximum communication paths & 6   \\
$T$ & Sensing targets & $\U\{0, \ldots, \Tmax\}$  \\
$\Ttilde$ & Communication paths & $\U\{1, \ldots, \Ttildemax\}$   \\
$\sigmarcst, \sigmarcstt$ & Target and scatterer \ac{RCS} & $\Exp(1/\sigmarcsmean)$  \\ 
$\sigmarcsmean$ & Mean \ac{RCS} & $1\ \mathrm{m}^2$   \\
$\measuredangle{(\alpha_t}), \measuredangle{(\alphatilde_t})$ & Phase of the channel gain & $\U[0,2\pi)$ \\
$\theta_t$ & Target angle & $\U[\thetamin, \thetamax]$ \\
$\thetatilde_t$ & \ac{UE} angle of departure & $\U[\thetatmin, \thetatmax]$   \\
$\thetamin, \thetatmin$ & Target and \ac{UE} minimum angle & $\thetamean - \Deltatheta/2$ \\
$\thetamax, \thetatmax$ & Target and \ac{UE} maximum angle & $\thetamean + \Deltatheta/2$ \\
$\thetamean$ & Mean angular uncertainty region & $\U[-60\degreee, 60\degreee]$ \\
$\Deltatheta$ & Angular deviation from $\thetamean$ & $\U[10\degreee, 20\degreee]$ \\
$R_t$ & Target range & $\U[\Rmin, \Rmax]$ \\
$[\Rmin, \Rmax]$ & Target range uncertainty region & $\U[10\ \text{m}, 43.75\ \text{m}]$ \\
$\Rtilde_1$ & \ac{UE} range & $\U[\Rtilde_{\min}, \Rtilde_{\max}]$  \\
$[\Rtilde_{\min}, \Rtilde_{\max}]$ & \ac{UE} range uncertainty region & $\U[10\ \text{m}, 200\ \text{m}]$ \\
$\SNRc$ & Communication SNR  & $21.1$ dB  \\ \midrule

$\mu$ & \multirow{2}{*}{GOSPA parameters in \eqref{eq:gospa_loss}} & $2$ \\
$p$ &  & $2$ \\
$B$ & Batch size & $4000$ samples \\
- & Training iterations & $10000$ \\
- & Learning rate & \begin{tabular}[c]{@{}c@{}}$10^{-2}$ for GPIs \\ $10^{-4}$ for ADIs\end{tabular}  \\ \midrule

- &  Testing samples  & $10^6$ \\
\bottomrule%
\end{tabular}%
}
\end{table}

\subsection{Evaluation Metrics} \label{subsec:eval_metric}
In this section we describe the metrics to evaluate the performance of the \ac{ISAC} system, which are:
\subsubsection{Misdetection Probability}
It refers to the probability that a target is missed during detection. In the case of multiple targets, the definition is adapted according to \cite{muth2023autoencoder}
\begin{align}
    \pmd = 1-\frac{\sum_{i=1}^B\min\{T_i, \That_i\}}{\sum_{i=1}^B T_i}. \label{eq:pmd_multi_targets}
\end{align}

\subsubsection{False Alarm Probability}
It refers to the probability that a measurement is incorrectly interpreted as a detected target. The definition is given by \cite{muth2023autoencoder}
\begin{align}
    \pfa = \frac{\sum_{i=1}^B (\max\{T_i, \That_i\} - T_i)}{\sum_{i=1}^B (\Tmax - T_i)}. \label{eq:pfa_multi_targets}
\end{align}

The termination threshold $\delta$ in Algorithm~\ref{alg:omp} determines the number of estimated targets, and hence, the misdetection and false alarm probabilities.

\subsubsection{\Ac{GOSPA}}
The \ac{GOSPA} loss~\cite{rahmathullah2017generalized} considers the number of estimated targets and their positions and it has been extensively applied in the literature~\cite{pinto2023deep, jones2023gospa, wang2024dynamic}. The \ac{GOSPA} loss is defined as follows. Let $\gamma>0$, $0<\mu\leq2$ and $1\leq p<\infty$. Let $\Pset = \{\bmp_1, \ldots, \bmp_{|\Pset|}\}$ and $\Psethat=\{ \bmphat_1, \ldots, \bmphat_{|\Psethat|} \}$ be the finite subsets of $\realset[2]$ corresponding to the true and estimated target positions, respectively. Let $d(\bmp, \bmphat)=\norm{\bmp-\bmphat}$ be the distance between true and estimated positions, and $\dcut(\bmp, \bmphat) = \min(d(\bmp, \bmphat),\gamma)$, where $\gamma$ is the cut-off distance. Let $\Pi_n$ be the set of all permutations of $\{1, \ldots, n\}$ for any $n\in\mathbb{N}$ and any element $\pi\in\Pi_n$ be a sequence $(\pi(1), \ldots, \pi(n))$. For $|\Pset|\leq|\Psethat|$, the GOSPA loss function is defined as 
\begin{align}
    &\mathcal{J}_p^{(\gamma,\mu)}(\Pset, \Psethat) = \nonumber \\
    &\bigg( \min_{\pi\in \Pi_{|\Psethat|}} \sum_{i=1}^{|\Pset|} \dcut(\bmp_i, \bmphat_{\pi(i)})^p + \frac{\gamma^p}{\mu}  (\abs{\Psethat}-\abs{\Pset}) \bigg)^{\frac{1}{p}}.
    \label{eq:gospa_loss}
\end{align}
If $\abs{\Pset} > \abs{\Psethat}, \mathcal{J}_p^{(\gamma,\mu)}(\Pset, \Psethat) = \mathcal{J}_p^{(\gamma,\mu)}(\Psethat, \Pset)$. As $p$ increases, the penalization applied to estimates far from the ground-truth targets becomes more severe. The value of $\gamma$ dictates the maximum allowable distance error. The role of $\mu$, together with $\gamma$, is to control the detection penalization. In the GOSPA loss of \eqref{eq:gospa_loss}, we set $\mu=2$ as recommended in \cite{rahmathullah2017generalized} and $\gamma=(\Rmax-\Rmin)=33.75$ m, which corresponds to the maximum range error.

\subsubsection{\Ac{SER}}
For communications, we measure the error between the transmitted symbols $\bmx$ and the estimated symbols $\hat{\bmx}$ by the \ac{SER}, defined as 
\begin{align}
    \mathrm{SER}=1/S\sum_{s=1}^S\Pr([\bmx]_s \neq [\hat{\bmx}]_s).
\end{align}
The \ac{SER} measures the average probability that the estimated symbol is not equal to the true transmitted symbol.

\subsection{Baselines}
To assess the performance of our proposed method, we compare it to the following baselines.

\subsubsection{Model-Based}
We consider a conventional model-based approach to compare with the proposed data-driven solution. The \ac{TX} is computed according to \eqref{eq:isac_precoder}, the sensing \ac{RX} follows the OMP Algorithm~\ref{alg:omp}, and the communication \ac{RX} estimates the symbols according to \eqref{eq:ml_detection}. We consider two cases: (i) the system has knowledge of the impairments, i.e., $\bmPsi=\bmXi$ and (ii) the system does not have knowledge of the impairments, i.e., we assume that the inter-antenna spacing is $\lambda/2$, i.e., $\bmomegatx=\bmomegarx=[-(K-1)\lambda/4, \ldots, (K-1)\lambda/4]^{\top}$ and no GPIs, i.e., $\bmxitx=\bmxirx=\bm{1}$. 

\subsubsection{\Ac{SLCB}}
In supervised learning, we assume that labeled data about the true target positions and communication symbols are available at the sensing and communication RXs, respectively. 
We modify the definition of the loss function in \eqref{eq:isac_opt_problem} as follows.
As sensing loss function, we adopt the loss of~\cite[Eq. (15)]{chatelier2025physically} for our \ac{ISAC} system. We consider the negative value of the \ac{ADM} evaluated at the true angle and delay of the targets, i.e., 
\begin{align} \label{eq:slcb_sens_loss}
    \Lr(\bmPsi) = - \frac{1}{T} \sum_{t=1}^T \lvert \bmarx^{\hermit}(\theta_t;\bmPsirx) {\Ys(\bmPsitx)} [\bmx \odot \bmrho(\tau_t)]^* \rvert^2.
\end{align}

For communications, we consider the loss function used in \cite{mateos2025model}, which leverages the \ac{CCE} loss based on an estimate of a probability vector of the true transmitted symbol on each subcarrier and the posterior distribution of the symbols. In our case, the CCE loss is expressed as
\begin{align}
    \Lc(\bmPsitx) = -\sum_{i=1}^{|\Xset|} [\bmxenc]_i \log[\bmchihat(\bmPsitx)]_i,
\end{align}
where $\bmxenc\in\{0,1\}^{|\Xset|}$ is the one-hot encoding vector corresponding to $[\bmx]_i$ and $\bmchihat(\bmPsitx)$ is the estimated posterior distribution of the symbols, computed as
\begin{align}
    \bmchihat(\bmPsitx) = \mathrm{Softmax}(-\log|[\yc(\bmPsitx)]_s - [\bmkappa(\bmPsitx)]_s \bmxref|^2),
\end{align}
with $\bmxref\in\complexset[|\Xset|]$ the vector containing all possible transmitted symbols. 
In the case of \ac{SLCB}, we consider that the true gradient of the channel function is known.

\subsection{Sensing Results} \label{subsec:sensing_results}
\subsubsection{Effect of the Loss Function} \label{subsubsec:results_roc}
First, we compare the performance of the loss functions in \eqref{eq:loss_r_1} and \eqref{eq:loss_r_2} when calibrating the \ac{RX} impairments. In this case, we consider that the impairments at the \ac{TX} are known and that there is no clutter ($\sigmac^2=0$), to focus on the effect of the sensing loss function and disregard the hyper-parameter selection of the \ac{GF} approach of Sec.~\ref{subsec:gf_backprop}; the effect of the clutter level is investigated separately in
Sec.~\ref{subsubsec:results_clutter}. 
We also assume that the \ac{ISAC} \ac{BS} illuminates both targets and the \ac{UE} based on higher-level protocols depending on the specific \ac{ISAC} application, which we model as $\omegar\sim\U[0,1]$ at each transmission. Given that we only need to calibrate the \ac{RX} impairments, we choose $\etar=1$. 

\begin{figure}
    \centering
    \includestandalone[width=0.47\textwidth]{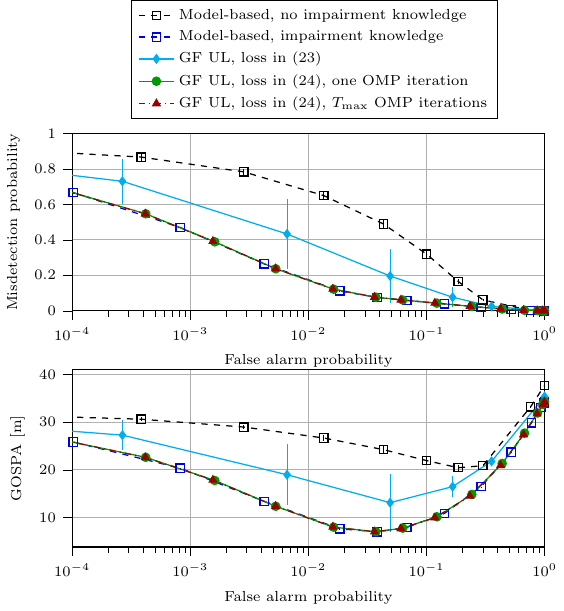}
    \caption{Sensing results as a function of the false alarm probability for five random realizations of the impairments and $\SCNRs=-3$ dB ($\sigmac^2=0$). The curves represent the mean performance over the impairment realizations and the error bars the standard deviation.}
    \label{fig:results_sensing_roc}
\end{figure}

In Fig.~\ref{fig:results_sensing_roc}, the sensing performance as a function of the false alarm probability is shown for the model-based method and the proposed \ac{GF} \ac{UL} approach. From Fig.~\ref{fig:results_sensing_roc}, it is observed that the loss in~\eqref{eq:loss_r_1} performs poorly on average and close to the model-based baseline with no impairment knowledge. The advantage of the loss in~\eqref{eq:loss_r_1} is a reduced complexity, which was shown to work in simpler scenarios with only one sensing target~\cite{mateos2025unsupervised}.
On the other hand, the loss in~\eqref{eq:loss_r_2} has a similar performance to the model-based baseline with known impairments. 
Moreover, considering one or $\Tmax$ \ac{OMP} iterations in~\eqref{eq:loss_r_2} does not produce significant changes in sensing performance. This suggests that removing the strongest target in the first iteration of the \ac{OMP} algorithm already indicates if the impairments are matched.
Consequently, weaker or overlapping targets do not compromise the proposed approach as long as the strongest reflection is received at a comparable \ac{SCNR}.
If the RX is uncalibrated, the mismatch between the estimated and true positions of the
first (and strongest) target creates spurious sidelobes in the \ac{ADM} after updating the
residual in~\eqref{eq:update_residual}, which increases the loss in~\eqref{eq:loss_r_2}, even when
other weaker targets are present. However, when the RX ULA is calibrated, the true and estimated
positions coincide (up to an error equal to the angular grid resolution) and the spurious sidelobes are greatly reduced. In the remainder of the paper, we will use the proposed \ac{GF} \ac{UL} loss in~\eqref{eq:loss_r_2} with one \ac{OMP} iteration.

\subsubsection{Effect of Clutter}
\label{subsubsec:results_clutter}
\begin{figure}
    \centering
    \includestandalone[width=0.47\textwidth]{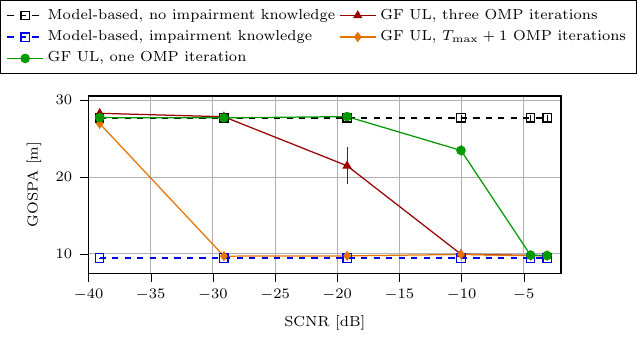}
    \caption{GOSPA as a function of the SCNR for a fixed false alarm probability of $10^{-2}$ and five random realizations of the impairments. The curves represent the mean performance over the impairment realizations and the error bars the standard deviation.}
    \label{fig:results_sensing_clutter}
\end{figure}

Given that the loss in~\eqref{eq:loss_r_2} greatly outperforms the loss in~\eqref{eq:loss_r_1}, we only consider in this section the loss in~\eqref{eq:loss_r_2}. In Fig.~\ref{fig:results_sensing_clutter}, we show the \ac{GOSPA} loss as a function of the \ac{SCNR}, where only $\sigmac^2$ is changed for different values of \ac{SCNR}. The results of Fig.~\ref{fig:results_sensing_clutter} do not consider clutter at testing so that the results are comparable across \ac{SCNR} levels and performance degradation comes from the calibration stage rather than from stronger clutter.
Note that clutter and targets are not distinguished by the proposed loss (both are coherent reflections observed through the same impaired arrays) so that clutter scatterers also carry information about the impairments.
In Fig.~\ref{fig:results_sensing_clutter}, it is observed that with a single \ac{OMP} iteration, the calibration performance remains close to that obtained with
known impairments for moderate clutter levels. Once the \ac{SCNR} falls below $-10$~dB, however,
a single iteration removes only the strongest reflection, and the residual in~\eqref{eq:loss_r_2}
becomes dominated by the remaining clutter, so that it is no longer informative about the
impairments.  As the number of OMP iterations increases, the proposed algorithm becomes more robust to the clutter power, which indicates that the clutter scatterers are canceled by the OMP algorithm and only the effect of targets remains, corresponding to the previous case of Fig.~\ref{fig:results_sensing_roc}. Nevertheless, introducing more iterations in the OMP algorithm increases the computational complexity of the calibration algorithm, as pointed out in Remark~\ref{remark:complexity}.

Since the clutter effect can be removed by clutter suppression techniques and more OMP iterations, in the remainder of the text, we consider that there are no clutter scatterers ($\sigmac^2=0$), and we consider a sensing \ac{SNR} of $-3$~dB unless stated otherwise.

\subsection{ISAC Results} \label{subsec:isac_results}
In this case, we consider both \ac{TX} and \ac{RX} parameters $\bmPsi$ and optimize them according to \eqref{eq:isac_opt_problem}. During training, we consider that the \ac{ISAC} \ac{BS} illuminates both targets and the \ac{UE} such that $\omegar\sim\U[0,1]$. During testing, we sweep over the values of $\omegar$ to obtain \ac{ISAC} trade-off curves.

\subsubsection{Effect of the Tradeoff Parameter $\etar$}
\begin{figure}
    \centering
    \includestandalone[width=0.47\textwidth]{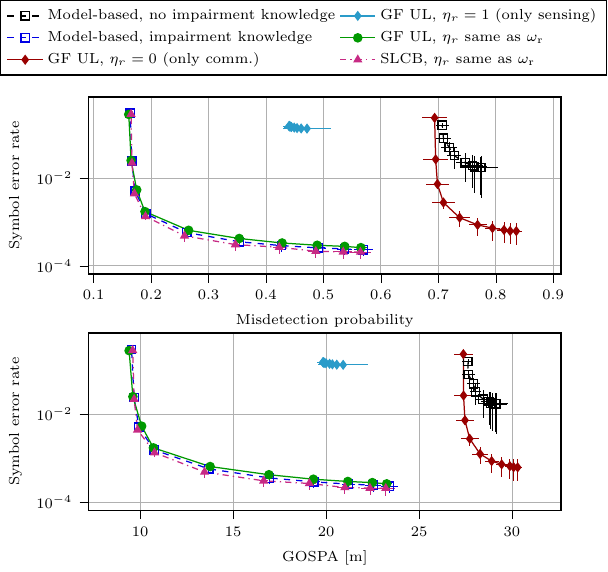}
    \caption{ISAC results for five realizations of the hardware impairments, a false alarm probability of $10^{-2}$, and $\SNRc = 21.1$ dB during inference. The curves represent the mean performance over the impairment realizations and the error bars the standard deviation.}
    \label{fig:results_isac}
\end{figure}

In Fig.~\ref{fig:results_isac}, we represent the inference \ac{ISAC} results over five random realizations of the impairments. We first consider training only using the communication ($\etar=0$) or sensing ($\etar=1$) losses. In the case of $\etar=0$, the communication performance is comparable to the baseline with known impairments. However, $\etar=0$ offers a poor sensing performance compared to the baseline with known impairments. As expected, the \ac{TX} parameters converge to a good solution, but the \ac{RX} parameters are not optimized because they are a function of the sensing loss. This case slightly outperforms the sensing performance of the model-based approach with no impairment knowledge because optimized \ac{TX} parameters close to the true impairments yield a better precoder and \ac{SNR} for sensing, as shown in Fig.~\ref{fig:example_bf}. 

The case of $\etar=1$ offers a poor communication performance and an improved sensing performance compared to $\etar=0$, but still worse performance than the model-based baseline with known impairments. This indicates that although both \ac{TX} and \ac{RX} parameters are now optimized, the sensing loss does not provide a good \ac{TX} parameter solution as the communication performance of $\etar=1$ is poorer than the model-based baseline with no impairment knowledge. The deficient solution of the \ac{TX} parameters implies a reduced sensing \ac{SNR} compared to matching the true \ac{TX} impairments, as shown in Fig.~\ref{fig:omp_example}, which explains the gap in the sensing performance between the model-based baseline with known impairments and the proposed approach with $\etar=1$.

Finally, when we let $\etar$ have the same realizations as $\omegar$, the \ac{ISAC} performance is close to known impairments and to \ac{SLCB}.
This suggests that training for sensing and communication effectively calibrates both \ac{TX} and \ac{RX} impairments. 
Moreover, \ac{SLCB} slightly outperforms the model-based baseline, which indicates that as the precoder function in~\eqref{eq:tx_proposed} is not optimal, the learned parameters $\bmPsitx$ yield a precoder that performs slightly better than~\eqref{eq:tx_proposed}.
In summary, the proposed \ac{GF} \ac{UL} approach yields an \ac{ISAC} performance similar to knowledge of the impairments when we combine sensing and communication objectives.

\subsubsection{Effect of Joint TX and RX Optimization} \label{subsubsec:results_joint_opt}

We now assess the benefit of optimizing the TX and RX parameters jointly, as opposed to
optimizing them individually in an alternate manner. In the latter case, the TX parameters are
first optimized from communication signals ($\etar=0$) and the RX parameters are subsequently
optimized from sensing signals ($\etar=1$), or vice versa. Hence, only one of the two parameter
sets is updated at each training step, and the sensing loss no longer contributes to the
optimization of the TX parameters.

\begin{figure}
    \centering
    \includestandalone[width=0.47\textwidth]{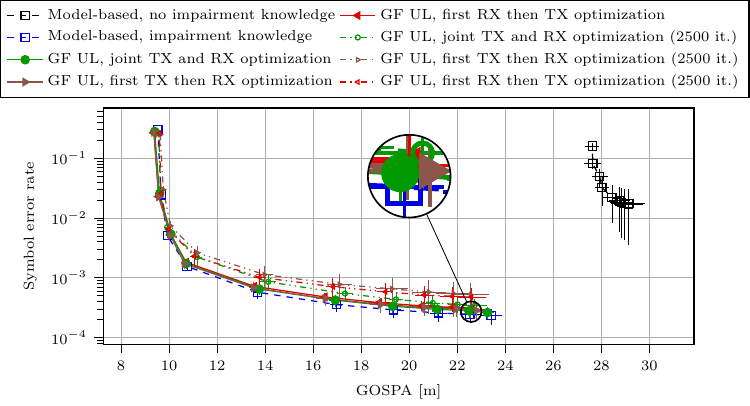}
    \caption{Symbol error rate as a function of GOSPA for different optimization strategies, five realizations of the hardware impairments, a false alarm probability of $10^{-2}$, and $\SNRc = 21.1$ dB during inference. The curves represent the mean performance over the impairment realizations and the error bars the standard deviation. (\textit{it.:} iterations.)}
    \label{fig:results_isac_joint_opt}
\end{figure}

Fig.~\ref{fig:results_isac_joint_opt} compares both strategies for the nominal number of training
iterations of Table~\ref{tab:sim_parameters} and for a reduced number of $2500$ iterations. With
the nominal number of iterations, alternate optimization attains an \ac{ISAC} Pareto front similar
to that of joint optimization, i.e., both strategies eventually converge to comparable operating
points. The two strategies differ, however, in how many iterations they need to reach them. When
the number of iterations is reduced to $2500$, all calibration strategies degrade, but joint
optimization degrades the least and remains the closest to the model-based benchmark with
impairment knowledge. 

Two effects explain this faster convergence. First, joint optimization updates the TX and the RX
parameters at every training step, whereas alternate optimization devotes only a fraction of the
available iterations to each parameter set. Second, the sensing loss also carries information
about the TX impairments and therefore contributes to the calibration of the TX parameters in the
joint formulation, while in the alternate one the TX parameters are calibrated from the
communication loss alone. In addition, when the RX is optimized first, its calibration is
performed under an uncalibrated precoder and hence under a reduced sensing \ac{SNR}.

These observations confirm that the TX and RX impairments, although physically independent, are
coupled through the end-to-end \ac{ISAC} performance. Exploiting this coupling does not change the
ultimately attainable \ac{ISAC} Pareto front, but how fast it is approached, as joint optimization
reaches it with fewer training iterations. It moreover avoids the design choices that
alternate optimization entails, namely the optimization order of the two parameter sets and the
switching iteration between them, which are not known a priori. Both aspects are relevant in our
setting, where calibration is performed online while the \ac{ISAC} \ac{BS} serves the \ac{UE} and
senses the environment: a faster convergence yields a better \ac{ISAC} service earlier in time,
and fewer design choices simplify the deployment of the calibration algorithm.

\subsection{Precoder Results}
\begin{figure}[tb]
    \centering 
     \includestandalone[width=0.47\textwidth]{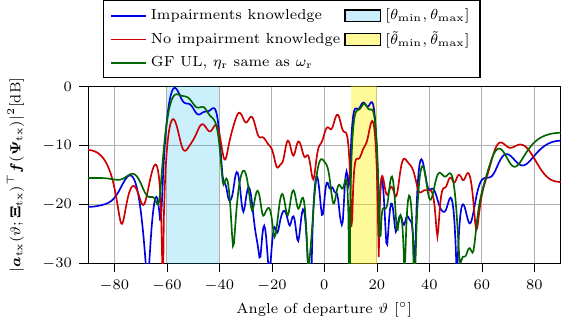}
    \caption{Precoder response $|\bmatx(\vartheta;\bmXitx)^\top \bmf(\bmPsitx)|^2$  for a sensing angular sector $[\thetamin, \thetamax] = [-60\degreee, -40\degreee]$ and a communication angular sector $[\thetatmin, \thetatmax] = [10\degreee, 20\degreee]$. The parameters $P$ and $\omegar$ in \eqref{eq:isac_precoder} are $P=0.1$ W, $\omegar=0.6$.}
    \label{fig:results_precoder}
\end{figure}

Under the same considerations of Sec.~\ref{subsec:isac_results}, Fig.~\ref{fig:results_precoder} shows the precoder response $|\bmatx(\vartheta;\bmXitx)^{\top}\bmf(\bmPsitx)|^2$ as a function of the angle of departure $\vartheta$ for one of the realizations of the impairments. Compared to the example in Fig.~\ref{fig:example_bf}, we include the precoder response with the optimized parameters $\bmPsitx$ of the proposed approach. The results in Fig.~\ref{fig:results_precoder} indicate that the learned parameters generate a precoder with a similar response to the case when the impairments are known ($\bmPsitx = \bmXitx$). This observation is consistent with the \ac{ISAC} results in Fig.~\ref{fig:results_isac}. Namely, the learned parameters yield a communication \ac{SNR} similar to knowledge of the impairments, implying a similar communication performance (the impairments affect the received communication signal $\yc$ in~\eqref{eq:comm_model} through $\bmatx^{\top}(\thetatilde_t)\bmf$) and increasing the likelihood of correct target estimation compared to no knowledge of the impairments ($\bmPsitx = [\bm{1}, \bmpideal]$). 

\subsection{Effect of the Precoder Perturbation} \label{subsec:prec_pert}
\begin{figure}[tb]
    \centering
    \includestandalone[width=0.49\textwidth]{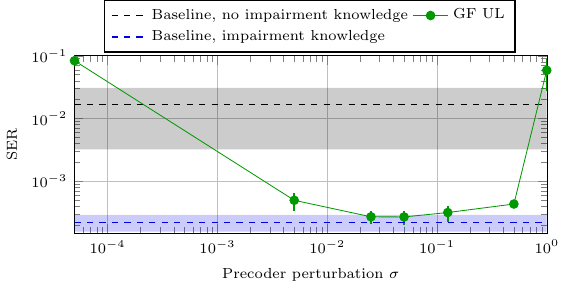}
    \caption{Symbol error rate as a function of the standard deviation of the precoder perturbation for $\SNRc = 21.1$ dB during inference. The curves represent the mean performance over the five impairment realizations, and the error bars and shaded areas the standard deviation.}
    \label{fig:results_prec_pert}
\end{figure}

We now investigate the impact of the precoder perturbation on the communication performance. We
follow the same training procedure as in Sec.~\ref{subsec:isac_results}, i.e., both the TX and the RX parameters are optimized. Fig.~\ref{fig:results_prec_pert} shows the SER as a function of the standard
deviation $\sigma$ of the precoder perturbation. The choice of $\sigma$ exhibits a tradeoff. Small values
of $\sigma$ hinder the exploration of precoders that achieve a lower communication loss, whereas large
values cause the approximate gradient in~\eqref{eq:grad_free} to decay rapidly, so that the optimization
may require significantly more training iterations. The range $\sigma\in[0.025,0.125]$ performs close to
perfect knowledge of the impairments, and the selected value $\sigma=0.05$ yields the best results
on average for the considered ISAC scenario. Hence, although hyper-parameter tuning is important to
approach the performance attained with perfect knowledge of the impairments, the gradient approximation
is robust to a wide range of perturbations.

\subsection{Effect of Angular Prior Sector} \label{subsec:effect_angular_sector}
So far, the calibration of the \ac{ISAC} \ac{BS} has assumed perfect prior knowledge of the angular sector in which the targets and the \ac{UE} lie.
This sector is obtained from an initial beam sweeping stage based on power measurements~\cite{xiao2024simultaneous}, which are robust but not immune to array calibration errors.
Hence, we assess here the robustness of the proposed calibration algorithm to such errors, which affect both the beamforming operation in~\eqref{eq:tx_proposed} and the construction of the angular-domain dictionary for \ac{OMP} in~\eqref{eq:angle_dict}.

We consider that the targets and the \ac{UE} lie in $[\thetamin, \thetamax]$ and $[\thetatmin, \thetatmax]$, respectively, and that nothing else lies outside those intervals. The a priori angular sector, however, is a perturbed version of the true one. We add \ac{AWGN} to the true angular sector, i.e., $\thetap\sim\normal(\theta, \sigmatheta^2)$, where $\theta$ denotes any of the four angles that define the prior angular sectors. During inference, the angular sectors are not perturbed, so that the results remain comparable with those of the previous sections.

\begin{figure}[tb]
    \centering
    \includestandalone[width=0.49\textwidth]{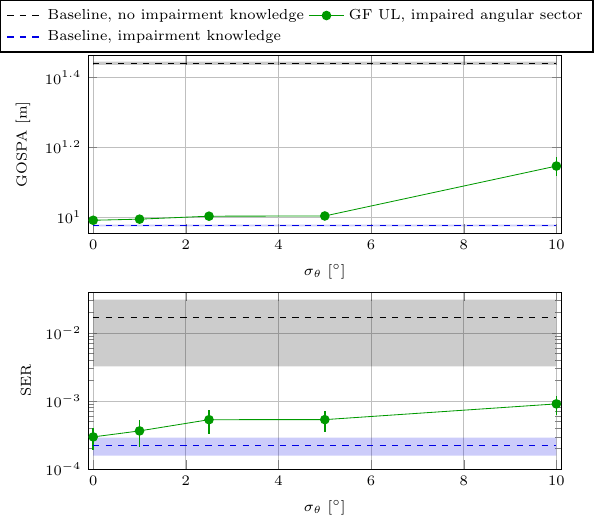}
    \caption{GOSPA (top) and symbol error rate (bottom) as a function of the angular sector perturbation. The curves represent the mean performance over the five impairment realizations, and the error bars and shaded areas the standard deviation.}
    \label{fig:results_sector_pert}
\end{figure}

Fig.~\ref{fig:results_sector_pert} shows the sensing and communication performance as a function of $\sigmatheta$. 
It is observed that sensing is more robust than communication, since the sensing performance is mainly driven by the \ac{RX} impairments, whereas the communication performance is affected solely by the \ac{TX} impairments, which are the ones estimated through the perturbed sector. 
In both cases, the performance degrades gradually with $\sigmatheta$, due to the reduced \ac{SNR} in the true angular region and the larger mismatch between the true and the perturbed angular sectors in the \ac{OMP} algorithm. 
Importantly, at the misalignment level reported after beam sweeping in~\cite{xiao2024simultaneous}  ($\sigmatheta\in[2.5,5]\degreee$), the degradation is minor.
Additionally, even for very severe misalignments ($\sigmatheta=10\degreee$), the proposed algorithm shows improved performance compared to the baseline with no impairment knowledge.

These results quantify the loss incurred by an imperfect angular prior and show that the proposed approach operates comfortably within the accuracy attainable by beam sweeping. In a practical deployment, the misalignment is further reduced as the \ac{ISAC} \ac{BS} is calibrated, so that performance improves over time at the potential cost of a longer calibration phase. Moreover, the sensing performance in a practical scenario may exceed that of Fig.~\ref{fig:results_sector_pert}, since clutter is likely to be present and can be exploited as an additional source of calibration.

\subsection{Generalization Tests} \label{results:generalization}
In this section, we evaluate the robustness of the proposed approach to a distribution shift, i.e., a mismatch between the training and the testing data. We consider two generalization tests that change: (i) the angular sector where the targets, UE, and scatterers appear and (ii) the sensing \ac{SNR}.

\subsubsection{Angular Sector Test} \label{subsubsec:generalization_angle}
\begin{figure}[tb]
    \centering
    \includestandalone[width=0.49\textwidth]{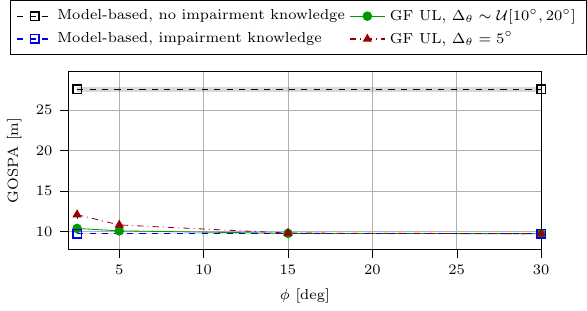}
    \caption{GOSPA as a function of the training angular sector. The curves represent the mean performance over the five impairment realizations, and the error bars and shaded areas the standard deviation.}
    \label{fig:results_generalization_angle}
\end{figure}

In this test, we modify the training angular sector for the targets $[\thetamin, \thetamax]$ and the UE together with the scatterers $[\thetatmin, \thetatmax]$. In particular, the mean angular sector is distributed according to $\thetamean\sim\U[-\phi, \phi]$. The testing angular sector remains as specified in Table~\ref{tab:sim_parameters}, i.e., for training the \ac{BS} \ac{FOV} is approximately $[-\phi, \phi]$ while for testing it is $[-60\degreee, 60\degreee]$. In Fig.~\ref{fig:results_generalization_angle}, we represent the GOSPA as a function of $\phi$. The model-based benchmarks show a flat behavior since the testing angular sector is unchanged. We observe that the proposed GF UL approach remains robust with a performance very close to knowledge of the impairments provided that the targets and the UE are not concentrated in boresight. If $\Deltatheta\sim\U[10\degreee, 20\degreee]$ the performance does not significantly change as a function of $\phi$, however, if the angular span is reduced to $\Deltatheta=5\degreee$, the GOSPA increases for lower values of $\phi$. This indicates that the effect of the \acp{ADI} is reduced when the ISAC transceiver illuminates boresight and it is therefore more challenging to compensate for \acp{ADI}.

\subsubsection{SNR Test}
\begin{figure}
    \centering
    \includestandalone[width=0.47\textwidth]{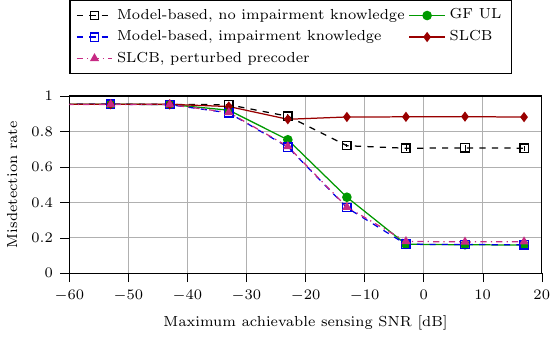}
    \caption{Misdetection probability as a function of the maximum achievable sensing SNR, when $\omegar\sim\U[0,1]$ and $\etar=\omegar$. The results consider five realizations of the impairments, where the points represent the mean performance and the error bars represent the standard deviation.}
    \label{fig:results_snr}
\end{figure}

Lastly, we test the generalization performance of the proposed approach and \ac{SLCB}. In particular, we reduce the training \ac{SNR} to $-33.0$ dB and we test the sensing performance for different sensing \acp{SNR}. Note that for lower \acp{SNR}, the effect of \ac{AWGN} is more pronounced than the effect of the impairments and array calibration becomes a more challenging problem. In Fig.~\ref{fig:results_snr}, we represent the misdetection probability as a function of the maximum achievable sensing \ac{SNR}. The maximum achievable sensing \ac{SNR} is chosen as a reference because the sensing \ac{SNR} at the \ac{RX} side depends on the \ac{TX} beamformer and is consequently affected by the \ac{TX} impairments. Fig.~\ref{fig:results_snr} shows that the performance of our proposed approach is similar to the baseline with known impairments, highlighting the effective calibration performance of the proposed method. However, the performance of \ac{SLCB} is far from the baseline with known impairments, which does not coincide with the \ac{ISAC} results of Fig.~\ref{fig:results_isac}. Our hypothesis is that the perturbation of the precoder $\bmfpert$ allows exploring more precoders, decreasing the likelihood of converging to a local minimum in the optimization problem of~\eqref{eq:isac_opt_problem}. To test this hypothesis, we include the results of \ac{SLCB} using the perturbed precoder $\bmfpert$ during training in the same manner as \ac{GF} \ac{UL}. In that case, the performance of \ac{SLCB} is much closer to the baseline with known impairments, which confirms our hypothesis. This result is similar to the effect of noise injection in \ac{DL}, which has also been shown to provide better generalization and convergence~\cite{jim1996analysis, nagabushan2016effect}. 

\subsection{Discussion on Practical Implementation} \label{subsec:implementation}
The proposed calibration algorithm is intended to run continuously online and the parameters are updated from the same transmissions that serve the sensing and communication functionalities. Therefore, no dedicated resources are reserved for calibration and the ISAC service is available from the first transmission, with a performance that improves as the calibration progresses. If the array response drifts (e.g., due to temperature changes), no periodic recalibration schedule is needed either: the proposed losses in~\eqref{eq:loss_r_2} and~\eqref{eq:loss_comm} require no labeled data and can therefore be monitored during normal operation at no additional cost, so that calibration is re-triggered whenever they degrade beyond a threshold.

On the other hand, the proposed
algorithm can be extended to multi-\ac{UE} communications. With digital arrays, the \acp{UE} are served at the same
time and frequency by superimposing one precoder per \ac{UE}~\cite{lai2004transmit, spencer2004zero}, so that
the composite transmitted signal, which remains known at the co-located sensing RX, replaces
$\bmf\bmx^{\top}$ in~\eqref{eq:sensing_model}. With analog arrays, the \acp{UE} are scheduled in
different time slots~\cite{gu2026optimal}. In both cases, the communication loss aggregates the feedback
of the scheduled \acp{UE} and the number of learnable parameters is unchanged, since they describe the array parameters.

\section{Conclusions} \label{sec:conclusions}
In this paper, we investigated the effect of \acp{GPI} and \acp{ADI} in an \ac{ISAC} \ac{BS}. We considered a scenario with multiple sensing targets and a \ac{UE} randomly distributed in the field-of-view of the \ac{ISAC} \ac{BS}. We first showed that under hardware impairments, the \ac{ISAC} precoder steers the energy in undesired directions and the response of the \ac{ADM} is significantly reduced and slightly shifted with respect to the true positions. 
We proposed a \ac{GF} \ac{UL} framework to calibrate the \ac{TX} and \ac{RX} impairments in the \ac{ISAC} \ac{BS}. 
Sensing results showed that minimizing the residual of the \ac{OMP} algorithm significantly outperforms maximizing the maximum response of the \ac{ADM}. Additionally, one iteration of the \ac{OMP} algorithm yields very similar results to using as many iterations as expected targets, reducing the computational complexity of the proposed approach.
\ac{ISAC} results showed that the proposed \ac{GF} \ac{UL} approach performs closely to \ac{SLCB} and to knowing the true impairments. 
Finally, we showed that the proposed approach generalizes better than \ac{SLCB} for a different testing sensing \ac{SNR} than during training due to the perturbation of the precoder needed to approximate the gradient of the channel.

Building on top of this work, promising research directions can consider: (i) calibration in non-line-of-sight scenarios, where modeling the propagation of the signals becomes more challenging and subject to mismatches; (ii) calibration under mutual coupling, where the ambiguities with GPIs must be resolved; (iii) reducing the computational complexity of the calibration algorithm via successive interference cancellation; or (iv) bistatic or multistatic scenarios, which can fuse the contributions of different BSs. Furthermore, experimentation with real hardware components can validate the effectiveness of the proposed \ac{GF} \ac{UL} approach.

\balance
\bibliographystyle{IEEEtran}
\bibliography{references}

\end{document}